\documentclass[journal]{IEEEtran}

\usepackage{graphicx}
\DeclareGraphicsExtensions{.png}

\usepackage[english]{babel}

\usepackage{savesym}
\savesymbol{AND}
\usepackage{algorithm,algorithmic}
\usepackage{setspace}
\usepackage{array}
\usepackage{caption}
\usepackage{url}
\usepackage{arydshln}

\usepackage{mathptmx}
\usepackage{amssymb}
\usepackage{amsfonts}
\usepackage{amsmath}
\usepackage{color}
\newtheorem{theorem}{Theorem}
\newtheorem{definition}{Definition}
\newtheorem{example}{Example}

\newtheorem{corollary}{Corollary}
\newtheorem{comment}{Remark}

\def\dem{\noindent {\it Proof:} }

\usepackage[most]{tcolorbox}

\usepackage{hyperref}
\usepackage{booktabs}
\usepackage{subfig}
\usepackage{xcolor}
\usepackage{amsmath}
\usepackage{graphicx}
\usepackage{tabularx}
\usepackage[
separate-uncertainty = true,
multi-part-units = repeat
]{siunitx}
\usepackage{supertabular,booktabs}
\usepackage{breqn}
\usepackage{adjustbox}

\newcommand{\notea}[1]{#1}

\newcommand{\noteb}[1]{#1}
\newcommand{\li}{L([0,1])}
\newcommand{\lr}{L(\mathbb{R})}
\newcommand{\note}[1]{\color{black}{#1} \color{black}}
\def\deme{\hfill $\Box$}

\newcommand\copyrighttext{%
	\footnotesize \textcopyright 2021 IEEE. Personal use of this material is permitted.
	Permission from IEEE must be obtained for all other uses, in any current or future
	media, including reprinting/republishing this material for advertising or promotional
	purposes, creating new collective works, for resale or redistribution to servers or
	lists, or reuse of any copyrighted component of this work in other works.
	DOI: \href{https://ieeexplore.ieee.org/document/9466935}{10.1109/TFUZZ.2021.3092824}}
\newcommand\copyrightnotice{%
	\begin{tikzpicture}[remember picture,overlay]
	\node[anchor=south,yshift=10pt] at (current page.south) {\fbox{\parbox{\dimexpr\textwidth-\fboxsep-\fboxrule\relax}{\copyrighttext}}};
	\end{tikzpicture}%
}
\begin{document}

\title{Interval-valued aggregation functions based on moderate deviations applied to Motor-Imagery-Based Brain Computer Interface}

\author{Javier Fumanal-Idocin, Zdenko Tak\'a\v{c}, Javier Fernández ~\IEEEmembership{Member,~IEEE}, Jose Antonio Sanz, Harkaitz Goyena, Ching-Teng Lin,~\IEEEmembership{Fellow,~IEEE},
	Yu-Kai Wang, ~\IEEEmembership{Member,~IEEE},
	Humberto Bustince,~\IEEEmembership{Senior,~IEEE}% <-this % stops a space
	
\thanks{Javier Fumanal-Idocin, Javier Fernandez, Jose Antonio Sanz, Harkaitz Goyena and Humberto Bustince are with the Departamento de Estadistica, Informatica y Matematicas, Universidad Publica de Navarra, Campus de Arrosadia, 31006, Pamplona, Spain.
	emails: javier.fumanal@unavarra.es, fcojavier.fernandez@unavarra.es , joseantonio.sanz@unavarra.es, bustince@unavarra.es}% <-this % stops a space
\thanks{Javier Fernandez and Humberto Bustince are with the Institute of Smart Cities, Universidad Publica de Navarra, Campus de Arrosadia, 31006, Pamplona, Spain.}
\thanks{Javier Fernandez and Humberto Bustince are with the Laboratory Navarrabiomed, Hospital Complex of Navarre (CHN), Universidad Publica de Navarra, IdiSNA, Irunlarrea 3. 31008, Pamplona, Spain.}
\thanks{Z.~Tak\'a\v{c} is with Institute of Information Engineering, Automation and Mathematics,
Faculty of Chemical and Food Technology, Slovak University of Technology in Bratislava, Radlinsk\'eho 9, 812 37 Bratislava 1, Slovak Republic,
(e-mail: zdenko.takac@stuba.sk).}% <-this % stops a space
\thanks{Y.-K. Wang and C.-T. Lin are with the Centre for Artificial Intelligence,
	Faculty of Engineering and Information Technology, University of Technology Sydney, Ultimo, NSW 2007, Australia (e-mail:, yukai.wang@uts.edu.au;
	chin-teng.lin@uts.edu.au)}
}
\markboth{IEEE Transactions on Fuzzy Systems}%
{Shell \MakeLowercase{\textit{et al.}}: Bare Demo of IEEEtran.cls for IEEE Journals}

\maketitle
\copyrightnotice
\begin{abstract}

\notea{In this work we develop moderate deviation functions to measure similarity and dissimilarity among a set of given interval-valued data to construct interval-valued aggregation functions, and we apply these functions in two Motor-Imagery Brain Computer Interface (MI-BCI) systems to classify electroencephalography signals. 
To do so, we introduce the notion of interval-valued moderate deviation function and, in particular, we study those interval-valued moderate deviation functions which preserve the width of the input intervals. In order to apply them in a MI-BCI system, we first use fuzzy implication operators to measure the uncertainty linked to the output of each classifier in the ensemble of the system, and then we perform the decision making phase using the new interval-valued aggregation functions. We have tested the goodness of our proposal in two MI-BCI frameworks, obtaining better results than those obtained using other numerical aggregation and interval-valued OWA operators, and obtaining competitive results versus some non aggregation-based frameworks.}
\end{abstract}

\section{Introduction}

\notea{Brain Computer Interface (BCI) is one of the most popular methods for controlling devices using variations in the brain dynamics \cite{lin2018wireless, lin2008noninvasive, lin2010real}. One popular BCI method is Motor-Imagery (MI), in which a person imagines a specific body movement, which produces a reaction in the motor areas of the brain \cite{park2012classification, wu2013common}. BCI systems are composed of \note{some} different components, such as signal detection, feature extraction and command identification, in order to successfully convert a brain signal into a computer command \cite{lotte2018review}. }

\notea{Usually, BCI systems use different wave transformations to extract useful information from the ElectroEncephaloGrahpy (EEG) data \cite{blankertz2011single, vaid2015eeg}, such as the Fast Fourier transform (FFt), to convert the signal in the frequency domain and the Meyer Wavelet transform. It is also very common to use algorithms such as Common Spatial Filtering (CSP), to classify the signals or to use its output as features to feed further classifiers \cite{wu2013common, blankertz2007optimizing}. Some of the most common classifiers used in BCI systems are Linear Discriminant Analysis \noteb{(LDA)}, Quadratic Discriminant Analysis \noteb{(QDA)}, Support Vector Machines \noteb{(SVM)} and K-Nearest Neighbours \noteb{(KNN)} \cite{alsaleh2016brain,vidaurre2010toward}.}

In the literature, there are many different approaches to EEG-based BCI classification systems. In \cite{Wu2017}, the authors extended the CSP to regression problems using fuzzy sets, and applied it to measure responsiveness in psychomotor vigilance tasks. In \cite{Jin2019}, the authors studied the correlation between different EEG channels and the target class, in order to select only meaningful channels to the classification problem, \noteb{while in \cite{jin2020internal} the authors purge the outliers from the signal and then use Dempster-Shafer theory to discover features with the highest interclass variability. In \cite{li2018effects} the authors studied the effects on visual stimuli, in order to understand how human perceive other people's emotions in the cocktail party problem \cite{cherry1953some}}. Also, in \cite{jin2020bispectrum}, the authors used Bispectrum analysis \cite{kotoky2014bispectrum} to select the optimal channels to perform classification.

\notea{One recent approach to BCI research is to focus on the information fusion processes of the system \cite{ko2019multimodal, wu2016fuzzy, fumanal2021motor}. Due to the high number of components of the BCI, it is necessary to combine the output from different elements into a single numerical value. This process is key to the performance of the system due to the relevance of these components interactions and correlations. One possibility to deal with this problem are aggregation functions \cite{beliakov2016practical, grabisch2009aggregation}.}

\notea{Aggregation functions are used to fuse several input values into one single output value. \note{They have been widely applied} in classification systems \cite{frank2007method, galar2011an}, fuzzy rules-based systems \cite{lucca2018cf, lucca2017cc} and image processing \cite{paternain2015construction, paternain2012color}, among others.}

\notea{In some cases, there is imprecision in the data to aggregate. \noteb{For the case of the EEG signal, the presence of noise and imprecision in the measurements can significantly affect the performance of the BCI system \cite{blankertz2011single}.} One solution to model that uncertainty is to represent each data as an interval, where its width represents the uncertainty associated to each observation \cite{historical, sanz2011genetic}. The use of intervals has shown to be a suitable solution to tackle classification problems \cite{sanz2014compact, sanz2012iivfdt, 9328158}. For this reason, large efforts have been devoted to the development of mechanisms to fuse information in the interval-valued setting \cite{entropyaccess,bustince2019similarity, takavc2019distances, ASMUS202027}.}

\notea{Taking into account these considerations, the objective of this paper is double:}
\begin{itemize}
	\item \notea{To construct a new MI BCI framework to classify EEG signals where the uncertainty in each classifier output is modeled using interval-valued data.}
	\item \notea{To determine the best aggregation function to be applied to the set of interval-valued data to obtain the final decision.}
\end{itemize}

\notea{The selection of the best aggregation function in an interval-valued setting is still an open problem. In the case of numerical data, several solutions have been proposed for this problem. In particular, in this work we consider the following ones: (i) \notea{Penalty-based aggregation functions, which determine the output from a set of inputs by minimizing a disagreement measure between the original set of values and the possible outputs \cite{bustince2017definition, bustince2014consensus}}. (ii) \notea{Deviation-based aggregation functions that were introduced in \cite{decky2018deviation} based on Darczy's deviation functions \cite{daroczy1972klasse}, which aggregate a set of deviation functions to determine how different is a given value from a set of inputs.} }

\notea{To reach our objective, we first develop the theoretical concept of interval-valued moderate deviation based aggregation function, studying the special case where the width of the input intervals is the same for all of them. Then, using the newly-developed interval-valued aggregation functions, we extend two aggregation-based MI BCI frameworks, namely: the traditional framework  described in \cite{wu2016fuzzy} and the Multimodal Fuzzy Fusion (MFF) framework proposed in \cite{ko2019multimodal}. }

The goodness of our proposal is shown by \noteb{comparing our results} \noteb{(i) with the outcome obtained by its numerical counterpart using classical aggregation functions, (ii) with the new method using as aggregation function the interval-valued OWA operators proposed in \cite{Bustince2013} and (iii), with other non aggregation-based MI BCI frameworks \cite{8553378, Lawhern2018, hbm23730}.}
	%: the work in \cite{8553378} where the authors extended the CSP and Riemannian covariance matrices to to multiscale and spectral cases, and \cite{lawhern2018eegnet}, where the authors used Convolutional Neural Networks to classify the EEG signals.}

%For these constructions, we are going to take into account both partial and total orders between intervals.

%To construct the intervals, we have used the methods proposed in \cite{Jurio2011}.
% In order to to show the effectiveness of these theoretical concepts, we have used these newly developed concepts in a Brain Computer Interface (BCI). }
%based on the aggregation-based systems in \cite{wu2016fuzzy, ko2019multimodal}

%Usually, to take a final decision an ensemble of classifiers is used, and the output is taken as the average of all their outputs. In \cite{ko2019multimodal}, the authors proposed the use of \note{aggregation functions, and, in particular, of different fuzzy integrals} to aggregate the classifiers output, outperforming the classical aggregations. We use the newly-developed aggregation functions to substitute the arithmetic mean in the traditional BCI framework \cite{wu2016fuzzy} and the fuzzy integrals in the BCI frameworks in \cite{ko2019multimodal}. The goodness of our proposal is shown by comparing it to standard numerical aggregations and the method based on interval-valued OWA operators proposed in \cite{Bustince2013}.

 \notea{The structure of this paper is described as follows. In Section \ref{sec:preliminary}, we explain some preliminary concepts related to the developed work. In Section \ref{sec:theory}, we discuss the interval-valued moderate deviations, and, in Section \ref{sec:theory2}, we discuss the specific case in which the length of all the interval-valued inputs  are the same. Then, in Section \ref{sec:md_bci}, we explain how to apply these functions in a MI BCI framework and, in Section \ref{sec:md_experiments}, we show our experimental results and comparisons with other aggregation  functions. \notea{Subsequently, in Section \ref{sec:cmp_others}, we show how our system performs compared to non aggregation-based MI BCI frameworks}. Finally, in Section \ref{sec:conclusions}, we summarize the work done and give the final remarks.}

\section{Preliminaries} \label{sec:preliminary}
In this section we introduce the concept of aggregation function, \notea{fuzzy implication function},  OWA operator and their interval-valued version.

\subsection{Aggregation functions}
\label{sec:agg}

Aggregation functions \noteb{\cite{grabisch2009aggregation}} are used to fuse information from $n$ sources into one single output. A function A: $[0, 1]^n \rightarrow [0,1]$ is said to be \notea{an} n-ary aggregation function if the following conditions hold:
\begin{itemize}
	\item A is increasing in each argument: \notea{$\forall x_i, y \in [0,1], i \in \{1,\ldots,n\}: x_i \le y \implies A(x_1, \ldots, x_i, \ldots, x_n) \le A(x_1, \ldots, y,  \ldots x_n)$}
	\item $A(0,\dots,0) = 0$
	\item $A(1,\dots,1) = 1$
\end{itemize}

Some examples of classical n-ary aggregation functions are:
\begin{itemize}
	\item Arithmetic mean: $A(\notea{\mathbf{x}}) = \frac{1}{n}\sum_{i=1}^{n} \notea{x}_i $;
	\item Median: $A(\notea{\mathbf{x}}) = \notea{x}_m$, where for any permutation $\sigma : \{1, \ldots, n\}$ such that  $\notea{x}_{\sigma(1)}\le \ldots \le \notea{x}_{\sigma(n)}, \notea{x}_m=\notea{x}_{\sigma(\frac{n+1}{2})}$, if $n$ is odd, and $\notea{x}_m = \frac{1}{2}(\notea{x}_{\sigma(\frac{n}{2})} + \notea{x}_{\sigma(\frac{n}{2}+1)})$, otherwise.
	\item Max: $A(\notea{\mathbf{x}}) = \max(\notea{x}_1, \dots, \notea{x}_n)$;
	\item Min: $A(\notea{\mathbf{x}}) = \min(\notea{x}_1, \dots, \notea{x}_n)$;
\end{itemize} \notea{where $\mathbf{x}=(x_1, \ldots, x_n) \in   [0,1]^n.$}

\subsection{\notea{Interval-valued fuzzy implication functions}}
An \notea{fuzzy implication function} is a function $I:[0,1]^2\rightarrow [0,1]$ that satisfies the following properties, \notea{for all $x, z, y\in[0,1]$}  \cite{bustince2003automorphisms}:
\begin{itemize}
  \item $x \le z$ implies $I(x,y) \ge I(z,y)$;
  \item $y\le t$ implies $I(x, y) \le I(x,t)$;
  \item $I(0,x) = 1$ for all $x \in [0,1]$;
  \item $I(x,1) = 1$ for all $x \in [0,1]$;
  \item $I(1,0) = 0$.
\end{itemize}
 \notea{Examples of fuzzy implication functions} are:
\begin{itemize}
  \item Kleene-Dienes: $I(x,y) = \max(1-x,y)$
  \item \L ukasiewicz: $I(x,y) = \min (1, 1- x + y)$
  \item Reichenbach: $I(x,y) = 1 - x + xy$
\end{itemize} \notea{ where $x,y  \in    [0,1].$}
%In order to create an interval from a given number in the unit interval, the following expression was proposed in \cite{Jurio2011}:
%\begin{equation}
%  F(x,y) = [\underline{I(x,y)}, \overline{I(x,y)+ y}]
%\end{equation}
%
%where $y \in [0,1]$.

%\subsection{Ordered Weighted Averaging operators (OWA) \cite{yager2012ordered}} \label{sec:owa}
%$\overrightarrow{w} = (w_1, ..., w_n) \in [0,1]^n$ is called a weighting vector if $\sum_{i=1}^{n} w_i = 1$. The OWA operator associated to $\overrightarrow{w}$ is the mapping OWA$_{\overrightarrow{w}}: [0, 1]^n \rightarrow [0,1]$ defined for every $(x_1, ..., x_n) \in [0,1]^n$ by:
%%They use use a quantifier function, $Q$, to generate a set of weights, $w$, to compute the following expression:
%
%\[
%OWA(x_1,...,x_n) = w_1x_{\gamma(1)} + ... + w_nx_{\gamma(n)}
%\]
%
%where $\gamma : \{1,...,n\} \rightarrow \{1,..,n\}$ is a permutation such that:
%$x_{\gamma(1)} \ge x_{\gamma(2)} \ge ... \ge x_{\gamma(n)}$.
%
%%Each $w_i$ is computed as:
%The weighting vector can be computed using a quantifier function, Q. For this study, we have used the following one:
%\[
%w_i = Q(\frac{i}{n}) - Q(\frac{i-1}{n})
%\]
%\[
%Q_{a,b}(i) =
%\begin{cases}
%0, & \mbox{if } i<a \\
%1, & \mbox{if } i>b \\
%\frac{i-a}{b-a}, & \mbox{otherwise}
%\end{cases}
%\]
%%
%where $a,b \in [0,1]$. Depending on the value of the parameters $a$ and $b$, different weighting vectors can be obtained. We have used three different ones:
%\begin{itemize}
%	\item OWA$_1$: $a=0.1, b=0.5$
%	\item OWA$_2$: $a=0.5, b=1$
%	\item OWA$_3$: $a=0.3, b=0.8$
%\end{itemize}

\subsection{Interval-valued aggregation functions \cite{bustince2013generation}} \label{sec:interval_aggs}
We consider closed subintervals of the unit interval $[0,1]$:
\begin{equation}
	L([0,1])=\{X=[\underline{X},\overline{X}]\mid 0\leq \underline{X}\leq \overline{X}\leq 1\}.
\end{equation}

The width of the interval $X\in L([0,1])$,  \notea{denoted by $w(X)$, is given by} $w(X)=\overline{X}-\underline{X}$. An \notea{interval-valued} function $f:(L([0,1]))^n\to L([0,1])$ is called $w$-preserving if $w(X_1)=\ldots=w(X_n)$ implies $w(f(X_1,\ldots,X_n))=w(X_1)$, \notea{$\text{for all } X_1, \ldots,X_n \in  L([0,1])$}.

An order relation on $L([0,1])$ is a binary relation $\leq$ on $L([0,1])$ such that, for all $X,Y,Z \in L([0,1])$,
\begin{enumerate}
\item[(L1)]
    $X\leq X$, ({\it reflexivity}),
\item[(L2)]
    $X\leq Y$ and $Y\leq X$ imply $X=Y$, ({\it antisymmetry}),
\item[(L3)]
    $X\leq Y$ and $Y\leq Z$ imply $X\leq Z$, ({\it transitivity}).
\end{enumerate}
An order relation on $L([0,1])$ is called total or linear if any two elements of $L([0,1])$ are comparable, i.e., if for every $X,Y \in L([0,1])$, $X \le Y$ or $Y \le X$.

We denote by $\leq_L$ any order in $L([0,1])$ (which can be partial or total) with $0_L=[0,0]$ as its minimal element and $1_L=[1,1]$ as its maximal element.

\notea{ The $K_a$ operator is defined, for all $X\in L([0,1])$ and $a \in [0,1]$, by:
\begin{equation}
	K_{a}(X) = (1-a)\underline{X}+a \overline{X}.
\end{equation}   For  $\alpha, \beta \in [0,1]$ with $\alpha\neq\beta$, the total order $\leq_{\alpha,\beta}$, induced by $K_\alpha$ and $K_\beta$, is defined, for all $X,Y\in L([0,1])$, as:} \cite{bustince2013generation}

\begin{equation}
	X\leq_{\alpha,\beta} Y \hspace{0.2cm} \text{ if } \hspace{0.2cm} \begin{cases}
	K_{\alpha}(X) < K_{\alpha}(Y) \text { or }
	\\
	K_{\alpha}(X)=K_{\alpha}(Y) \text{ and }
	K_{\beta}(X) \leq K_{\beta}(Y)
	\end{cases}
\end{equation}  \notea{A total order on $L([0,1])$ is called an admissible order  \cite{bustince2013generation,ASMUS202027} if it generalizes the standard product order \cite{moore2009,DIM00} between intervals, which is a partial order.}

\begin{definition}\label{def:ivagrfun}
	\notea{Consider} $n \ge 2$. An $n$-dimensional interval-valued aggregation function in $\li$ with respect to $\leq_{L}$ is a mapping $A:(\li)^n  \to \li$ which verifies:
	\begin{enumerate}
		\renewcommand{\labelenumi}{(\roman{enumi})}
		\item[(A1)]
		$A([0,0],\ldots,[0,0])=0_L$.
		\item[(A2)]
		$A([1,1],\ldots,[1,1])=1_L$.
		\item[(A3)]
		$A$ is a non-decreasing function in each variable with respect to $\leq_{L}$.
	\end{enumerate}
\end{definition}

\subsection{Interval-valued Ordered Weighted Averaging (OWA) operators} \label{sec:owa}
Let $\preceq$ be an admissible order on $L([0,1])$, \notea{$[a, b] \in L([0,1])$}  and $\mathbf{w} = (w_1, \dots, w_n) \in [0,1]^n, w_1 + \dots + w_n = 1$ a weighting vector. \notea{An} interval-valued OWA operator associated with $w$ and $\preceq$ is a mapping $L([0,1]^n \rightarrow L([0,1]))$ defined by \cite{Bustince2013}:
\begin{equation}
	OWA([a_1, b_1], \dots, [a_n, b_n]) = \sum_{i=1}^{n} w_i \cdot [a_{\sigma(i)}, b_{\sigma(i)}]
\end{equation}
where $\sigma$ is a permutation such that, for all \notea{$i \in \{1,\dots, n-1\}$, it holds that} $[a_{\sigma(i)}, b_{\sigma(i)}] \preceq [a_{\sigma(i+1)}, b_{\sigma(i+1)}]$.

The weighting vector can be computed using a quantifier function  \notea{$Q:[0,1] \rightarrow [0,1]$, defined here, for all $x \in [0,1]$ and $a,b \in [0,1]$, by: \begin{equation}
Q_{a,b}(x) =
\begin{cases}
0, & \mbox{if } x<a \\
1, & \mbox{if } x>b \\
\frac{x-a}{b-a}, & \mbox{otherwise}
\end{cases}
\end{equation}
We then define, for $i \in \{1 \dots n\}$}:
\begin{equation}
w_i = Q(\frac{i}{n}) - Q(\frac{i-1}{n})
\end{equation}
Depending on the value of the parameters $a$ and $b$, different weighting vectors can be obtained.  For this study, we have used the following \notea{configurations}:
\begin{itemize}
	\item OWA$_1$: $a=0.1, b=0.5$
	\item OWA$_2$: $a=0.5, b=1$
	\item OWA$_3$: $a=0.3, b=0.8$
\end{itemize}

\subsection{Moderate \noteb{d}eviation functions}

Regarding the notion of moderate deviation function, we follow the approach given in \cite{decky2017}.

\begin{definition}\label{def_ivmoddev}
A function $\mathcal{D}:[0,1]^2 \to \mathbb{R}$ is called a moderate deviation function, if, \notea{for all  $x,y \in [0,1]$},  it satisfies:
\begin{enumerate}
  \item[(MD1)] $\mathcal{D}$ is non-decreasing in the second component;
  \item[(MD2)] $\mathcal{D}$ is non-increasing in the first component;
  \item[(MD3)] $\mathcal{D}(x,y)=0$ if and only if $x=y$.
\end{enumerate}
The set of all moderate deviation functions is denoted by $MD$.
\end{definition}

The notion of moderate deviation function is closely related to those of restricted equivalence function~\cite{bustince2006restricted} that we recall now.

\begin{definition}\label{def:ref}
A function $R:[0,1]^2 \to [0,1]$ is called a restricted equivalence function, if, \notea{for all $x,y,z \in [0,1]$},  it satisfies:
\begin{enumerate}
\item[(R1)] $R(x,y)=0$ if and only if $\{x,y\}=\{0,1\}$;
\item[(R2)] $R(x,y)=1$ if and only if $x=y$;
\item[(R3)] $R(x,y)=R(y,x)$;
\item[(R4)] If $x\leq y\leq z$, then $R(x,z)\leq R(x,y)$ and $R(x,z)\leq R(y,z)$.
\end{enumerate}
\end{definition}

\section{Interval-valued moderate deviation functions} \label{sec:theory}

%We have applied the interval-valued moderate deviation functions in both the traditional BCI framework and the MFF BCI Framework.
A moderate deviation function was introduced and corresponding deviation-based aggregation functions were studied in \cite{decky2017}. We make a similar study for intervals.

\begin{definition}\label{def_ivmoddev_2}
Let $\leq_L$ be a total order on $\li$ and 
\begin{equation}
	\lr=\{A=[\underline{A},\overline{A}]\mid \underline{A}, \overline{A} \in \mathbb{R}, \underline{A} \leq \overline{A} \}
\end{equation}
A function $D:(\li)^2 \to \lr$ is called an interval-valued moderate deviation function w.r.t. $\leq_L$, if, \notea{for $X, Y \in L([0,1])$},  it satisfies:
\begin{enumerate}
  \item[(MD1)] $D$ is non-decreasing in the second component w.r.t. $\leq_L$;
  \item[(MD2)] $D$ is non-increasing in the first component w.r.t. $\leq_L$;
  \item[(MD3)] $D(X,Y)=0_L$ if and only if $X=Y$.
\end{enumerate}
The set of all interval-valued moderate deviation functions w.r.t. $\leq_L$ is denoted by $IMD$.
\end{definition}

Interval-valued aggregation functions based on a given moderate deviation function can be defined by: $A(X_1,\ldots,X_n)=Y$ if and only if the equation $\sum\limits_{i=1}^n D(X_i,Y)=[0,0]$ is satisfied for $X_1,\ldots,X_n\in\li$. However, from Definition~\ref{def_ivmoddev_2} it is clear that the equation may not have a solution, or it may have more than one solution. Hence, we modify the procedure in a similar way as it was done in \cite{decky2017}. We adopt the convention $\sup\{a\in [b,c]\,|\,a\in\emptyset\}=b$ and $\inf\{a\in [b,c]\,|\,a\in\emptyset\}=c$.

\begin{definition}\label{def_ivdmean}
Let $n\in\mathbb{N}$, $\leq_L$ be a total order on $\li$, $A:(\li)^2 \to \li$ be an idempotent interval-valued aggregation function w.r.t. $\leq_L$ and $D:(\li)^2 \to \lr$ be an interval-valued moderate deviation function w.r.t. $\leq_L$. Then the function $M_D: (\li)^n \to \li$ defined, \notea{for $X_1,\ldots,X_n \in L([0,1])$}, by

\begin{multline}
  M_D(X_1,\ldots,X_n)=A\left( \sup\left\{ Y\in\li \, \left|\, \sum\limits_{i=1}^n D(X_i,Y)<_L[0,0] \right. \right. \right\},  \\
  \inf\left\{ Y\in\li \,\left|\, \sum\limits_{i=1}^n D(X_i,Y)>_L[0,0] \right\} \right)
\end{multline}
is called an interval-valued $D$-mean w.r.t. $\leq_L$.
\end{definition}

We are going to show that the proposed $D$-mean is a symmetric idempotent interval-valued aggregation function.

\begin{theorem}\label{th_ivdmean}
Let $n\in\mathbb{N}$, $\leq_L$ be a total order on $\li$ and $D:(\li)^2 \to \lr$ be an interval-valued moderate deviation function w.r.t. $\leq_L$. Then the interval-valued $D$-mean $M_D: (\li)^n \to \li$ given in Definition~\ref{def_ivdmean} is an $n$-ary symmetric idempotent interval-valued aggregation function.
\end{theorem}
\dem
The symmetry is obvious. Regarding idempotency, let $X_1=\ldots=X_n$. Then, since $D(X_1,Y)<_L [0,0]$ if and only if $Y<_L X_1$, we have
\begin{equation}
	\sup\left\{ Y\in\li \,\left|\, \sum\limits_{i=1}^n D(X_i,Y)<_L[0,0] \right. \right\}=X_1
\end{equation}
and similarly
\begin{equation}
\inf\left\{ Y\in\li \,\left|\, \sum\limits_{i=1}^n D(X_i,Y)>_L[0,0] \right. \right\}=X_1,
\end{equation}
hence $M_D(X_1,\ldots,X_1)=A(X_1,X_1)=X_1$.

It suffices to prove that $M_D$ is an \noteb{interval-valued} aggregation function. The boundary conditions follow from the idempotency, so it only remains to show the monotonicity. Let $X_1,\ldots,X_n\in\li$ and let there exists $k\in\{1,\ldots,n\}$ such that $X_k<_L [1,1]$. Let $Z>_L X_k$. Then, since $D(Z,Y)\leq_L D(X_k,Y)$ for all $Y\in\li$, we have
$$
\sup\left\{ Y\in\li \,\left|\, \sum\limits_{i=1}^n D(X_i,Y)-D(X_k,Y)+D(Z,Y)<_L[0,0] \right. \right\} \geq_L
$$
\begin{equation}
\geq_L \sup\left\{ Y\in\li \,\left|\, \sum\limits_{i=1}^n D(X_i,Y)<_L[0,0] \right. \right\}
\end{equation}
and
$$
\inf\left\{ Y\in\li \,\left|\, \sum\limits_{i=1}^n D(X_i,Y)-D(X_k,Y)+D(Z,Y)>_L[0,0] \right. \right\} \geq_L
$$
\begin{equation}
\geq_L \inf\left\{ Y\in\li \,\left|\, \sum\limits_{i=1}^n D(X_i,Y)>_L[0,0] \right. \right\}.
\end{equation}
Having in mind the monotonicity of $A$, the monotonicity of $M_D$ is proved.
\deme

\begin{example}
Let $D:(\li)^2 \to \lr$ be given, \notea{for $X, Y \in L([0,1])$}, by
$$
D(X,Y)=\left\{
         \begin{array}{ll}
           Z, & \hbox{if } Y >_L X; \\
           -Z, & \hbox{if } Y <_L X; \\[0cm]
           [0,0], & \hbox{if } Y=X,
         \end{array}
       \right.
$$
for some $Z\in\li$ such that $Z >_L [0,0]$. Then, $D\in IMD$ and the corresponding \noteb{interval-valued} $D$-mean is median if $A$ is the median if $A$ is the arithmetic mean.

\end{example}

\begin{definition}\label{def_3dmeans}
Let $n\in\mathbb{N}$, $\leq_L$ be a total order on $\li$, $A:(\li)^2 \to \li$ be an idempotent interval-valued aggregation function w.r.t. $\leq_L$ and $\mathbb{D}=(D_1,\ldots,D_n)\in IMD^n$. Let $\mathbf{w}\in[0,\infty[^n$ be a non-zero weighting vector. Then
\begin{itemize}
  \item The function $M_{\mathbb{D}}: (\li)^n \to \li$ defined by
       $$
       M_{\mathbb{D}}(X_1,\ldots,X_n)=
       $$
       \begin{multline}
       =A\left( \sup\left\{ Y\in\li \, \left|\, \sum\limits_{i=1}^n D_i(X_i,Y)<_L[0,0] \right. \right\}, \right. \\
       \left. \inf\left\{ Y\in\li \,\left|\, \sum\limits_{i=1}^n D_i(X_i,Y)>_L[0,0] \right\} \right) \right.
		\end{multline}
       is called an interval-valued $\mathbb{D}$-mean w.r.t. $\leq_L$.
  \item The function $M_{\mathbb{D},\mathbf{w}}: (\li)^n \to \li$ defined by
       $$
       M_{\mathbb{D},\mathbf{w}}(X_1,\ldots,X_n)=
       $$
        \begin{multline}
       =A\left( \sup\left\{ Y\in\li \,\left|\, \sum\limits_{i=1}^n w_iD_i(X_i,Y)<_L[0,0] \right. \right\} \right., \\
       \left. \inf\left\{ Y\in\li \,\left|\, \sum\limits_{i=1}^n w_iD_i(X_i,Y)>_L[0,0] \right. \right\} \right),
        \end{multline}
       is called an interval-valued weighted $\mathbb{D}$-mean w.r.t. $\leq_L$.
  \item The function $OM_{\mathbb{D},\mathbf{w}}: (\li)^n \to \li$ defined by
       $$
       OM_{\mathbb{D},\mathbf{w}}(X_1,\ldots,X_n)=
       $$
       \begin{multline}
       =A\left( \sup\left\{ Y\in\li \,\left|\, \sum\limits_{i=1}^n w_iD_i(X_{(i)},Y)<_L[0,0] \right. \right\}, \right.\\
       \left. \inf\left\{ Y\in\li \,\left|\, \sum\limits_{i=1}^n w_iD_i(X_{(i)},Y)>_L[0,0] \right. \right\} \right),
        \end{multline}
      where $X_{(1)}\geq_L \ldots \geq_L X_{(n)}$, is called an interval-valued ordered weighted $\mathbb{D}$-mean w.r.t. $\leq_L$.
\end{itemize}
\end{definition}

\begin{corollary}
Under the conditions of Definition~\ref{def_3dmeans}, the interval-valued $\mathbb{D}$-mean $M_{\mathbb{D}}$, interval-valued weighted $\mathbb{D}$-mean $M_{\mathbb{D},\mathbf{w}}$ and interval-valued ordered weighted $\mathbb{D}$-mean $OM_{\mathbb{D},\mathbf{w}}$ are idempotent interval-valued aggregation functions. Moreover, $OM_{\mathbb{D},\mathbf{w}}$ is symmetric.
\end{corollary}

\section{$w$-Preserving interval-valued moderate deviation function} \label{sec:theory2}

In this section, a special class of \noteb{interval-valued} moderate deviation functions is studied, in particular, the functions satisfying the property: if all the input intervals have the same width, then the output interval has also the same width. We only consider $K_{\alpha,\beta}$ orders in this section.

\begin{definition}\label{def_ivwmoddev}
Let $\alpha,\beta\in[0,1]$ with $\beta\neq\alpha$. A function $D:(\li)^2 \to \lr$ is called a $w$-preserving interval-valued moderate deviation function w.r.t. $\leq_{\alpha,\beta}$, if,  \notea{for $X, Y \in L([0,1])$}, it satisfies:
\begin{enumerate}
  \item[(MD1)] $D$ is non-decreasing in the second component w.r.t. $\leq_{\alpha,\beta}$;
  \item[(MD2)] $D$ is non-increasing in the first component w.r.t. $\leq_{\alpha,\beta}$;
  \item[(MD3$^{\prime}$)] $K_{\alpha}(D(X,Y))=0$ if and only if $K_{\alpha}(X)=K_{\alpha}(Y)$;
%  \item[(MD3$^{\prime}$)] $K_{\alpha}(D(X,Y))=0$ if and only if $X=Y$;
  \item[(MD4)] if $w(X)=w(Y)$, then $w(D(X,Y))=w(X)$.
\end{enumerate}
The set of all $w$-preserving interval-valued moderate deviation functions w.r.t. $\leq_{\alpha,\beta}$ is denoted by $wIMD$.
\end{definition}

A weaker form of monotone \noteb{interval-valued} functions will be of use later, so we define the so-called $w$-monotonicity.

\begin{definition}
An interval-valued function $A: (\li)^n \to \li$ is said to be $w$-monotone function w.r.t. order $\leq_{L}$, if it satisfies:
\begin{enumerate}
\renewcommand{\labelenumi}{(\roman{enumi})}
\item[(A3$^{\prime}$)] If $X_1\leq_{L} Y_1,\ldots,X_n\leq_{L} Y_n$ where $w(X_1)=\ldots=w(X_n)=w(Y_1)=\ldots=w(Y_n)$, then $A(X_1,\ldots,X_n)\leq_{L} A(Y_1,\ldots,Y_n)$.
\end{enumerate}
\end{definition}

\begin{definition}\label{def_ivwdmean}
Let $\alpha,\beta\in[0,1]$ with $\beta\neq\alpha$, $n\in\mathbb{N}$ and $B:[0,1]^n\to[0,1]$ be an idempotent symmetric aggregation function. Let $A:(\li)^2 \to \li$ be an idempotent interval-valued $w$-monotone function w.r.t. $\leq_{\alpha,\beta}$ and $D:(\li)^2 \to \lr$ be a $w$-preserving interval-valued moderate deviation function w.r.t. $\leq_{\alpha,\beta}$. Then the function $M_D: (\li)^n \to \li$ defined by

\begin{multline}\label{eq_wmd}
M_D(X_1,\dots,X_n)= \\
=A\left(  \sup\biggl\{ Y\in\li \left| \sum_{i=1}^{n}S K_{\alpha}\left(D(X_i,Y)\right) <0 \right. \biggr. \right.  \\
 \biggl. w(Y)=B(w(X_1),\dots,w(X_n)) \biggr\}  ,   \\
 \inf\left\{ Y\in\li \left| \sum\limits_{i=1}^n K_{\alpha}(D(X_i,Y) \right. \right. )>0 \\
\biggl. \biggl. w(Y)=B(w(X_1),\dots,w(X_n)) \biggr\}\biggl)
\end{multline}

is called an interval-valued $wD$-mean w.r.t. $\leq_{\alpha,\beta}$.
\end{definition}

The following theorem shows that the proposed $wD$-mean has the similar properties as $D$-mean defined in Definition~\ref{def_ivdmean}, i.e. it is a symmetric idempotent interval-valued function. But unlike $D$-mean, our $wD$-mean is also $w$-preserving (if $A$ is $w$-preserving); on the other side it does not satisfy monotonicity, only the $w$-monotonicity.

\begin{theorem}\label{th_wivwmonfun}
Let $\alpha,\beta\in[0,1]$ with $\beta\neq\alpha$, $n\in\mathbb{N}$ and $D:(\li)^2 \to \lr$ be a $w$-preserving interval-valued moderate deviation function w.r.t. $\leq_{\alpha,\beta}$. Then the interval-valued $wD$-mean $M_D: (\li)^n \to \li$ given in Definition~\ref{def_ivwdmean} where $A$ is $w$-preserving, is an $n$-ary $w$-preserving symmetric idempotent interval-valued $w$-monotone function.
\end{theorem}
\dem
The symmetry is obvious. Idempotency follows from the equivalence $K_{\alpha}(D(X,Y))<0$ if and only if $K_{\alpha}(Y)<K_{\alpha}(X)$ (recall that $w(Y)$ has to be equal to $w(X)$, hence the last inequality is equivalent to $Y<_{\alpha,\beta}X$) in a similar way as in Theorem~\ref{th_ivdmean}.

Regarding the $w$-monotonicity of the function $M_D$, let $X_1,\ldots,X_n\in\li$ with $w(X_1)=\ldots=w(X_n)$ and let there exists $k\in\{1,\ldots,n\}$ such that $X_k<_{\alpha,\beta} [1,1]$. Let $Z>_{\alpha,\beta} X_k$ where $w(Z)=w(X_k)$. Then, since $D(Z,Y)\leq_{\alpha,\beta} D(X_k,Y)$ for all $Y\in\li$, we have
\begin{multline}
\sup\left\{ Y\in\li \,\left|\, \sum\limits_{i=1}^n K_{\alpha}(D(X_i,Y))-K_{\alpha}(D(X_k,Y)) \right. \right.\\
\biggl. +K_{\alpha}(D(Z,Y))<0 \,\,\, \& \,\,\, w(Y)=w(X_1) \biggr\} \geq_{\alpha,\beta} \\
\geq_{\alpha,\beta} \sup\left\{ Y\in\li \,\left|\, \sum\limits_{i=1}^n K_{\alpha}(D(X_i,Y))<0 \,\,\, \& \,\,\, w(Y)=w(X_1) \right. \right\}
\end{multline}
and
\begin{multline}
\inf\left\{ Y\in\li \,\left|\, \sum\limits_{i=1}^n K_{\alpha}(D(X_i,Y))-K_{\alpha}(D(X_k,Y))+ \right. \right.\\
\left. K_{\alpha}(D(Z,Y))>0 \,\,\, \& \,\,\, w(Y)=w(X_1) \right\} \geq_{\alpha,\beta}
\geq_{\alpha,\beta} \inf\left\{ Y\in\li \,\left|\, \right. \right.\\
\left. \sum\limits_{i=1}^n K_{\alpha}(D(X_i,Y))>0 \,\,\, \& \,\,\, w(Y)=w(X_1) \right\}
\end{multline}
Having in mind the $w$-monotonicity of $A$, the $w$-monotonicity of $M_D$ is proved.

Finally, the fact that $M_D$ is $w$-preserving directly follows from Equation~\eqref{eq_wmd}, idempotency of $B$ and the fact that A is w-preserving.
\deme

A construction method of $w$-preserving interval-valued moderate deviation functions is given in the following Theorem.

\begin{theorem}\label{th_con1_iwmd}
Let $\alpha,\beta\in[0,1]$ with $\beta\neq\alpha$, $\mathcal{D}:[0,1]^2 \to \mathbb{R}$ be a strictly monotone moderate deviation function and $C:[0,1]^2 \to [0,1]$ be an idempotent function non-decreasing in the second component and non-increasing in the first component. Then the function $D:(\li)^2 \to \lr$ given by:
$$
       \left\{
         \begin{array}{l}
           K_{\alpha}(D(X,Y))=\mathcal{D}\left(K_{\alpha}(X),K_{\alpha}(Y)\right), \\
           w(D(X,Y))=C\left(w(X),w(Y)\right)
         \end{array}
       \right.
$$
is a $w$-preserving interval-valued moderate deviation function w.r.t. $\leq_{\alpha,\beta}$.
\end{theorem}
\dem
(MD1) Let $Y\leq_{\alpha,\beta} Z$. There are two possibilities:
\begin{enumerate}
  \item $K_{\alpha}(Y)<K_{\alpha}(Z)$, then, since $\mathcal{D}$ is strictly monotone, we have $K_{\alpha}(D(X,Y))<K_{\alpha}(D(X,Z))$, hence $D(X,Y)<_{\alpha,\beta}D(X,Z)$, or
  \item $K_{\alpha}(Y)=K_{\alpha}(Z)$ and $K_{\beta}(Y)\leq K_{\beta}(Z)$, in which case $K_{\alpha}(D(X,Y))=K_{\alpha}(D(X,Z))$ and
\begin{multline}
	\left\{
	\begin{array}{ll}
	w(D(X,Y))=C\left(w(X),w(Y)\right)\leq \\C\left(w(X),w(Z)\right)=w(D(X,Z)), & \textrm{ for } \beta>\alpha, \\
	w(D(X,Y))=C\left(w(X),w(Y)\right)\geq \\C\left(w(X),w(Z)\right)=w(D(X,Z)), & \textrm{ for } \beta<\alpha \\
	\end{array}
	\right.	
\end{multline}

so, in both cases $K_{\beta}(D(X,Y))\leq K_{\beta}(D(X,Z))$ and finally $D(X,Y)\leq_{\alpha,\beta}D(X,Z)$.
\end{enumerate}

(MD2) can be proved similarly to (MD1). (MD3$^{\prime}$) follows from the fact that $\mathcal{D}$ is a moderate deviation function and (MD4) immediately follows from the idempotency of $C$.
\deme

\begin{example}\label{ex_wimd}
\begin{enumerate}
  \item[(i)] Taking $\mathcal{D}:[0,1]^2 \to \lr$ defined by $\mathcal{D}(x,y)=y-x$ and $C:[0,1]^2 \to [0,1]$ defined by $C(x,y)=\max\left(0,\min(1,2y-x)\right)$, by Theorem~\ref{th_con1_iwmd} one obtains a $w$-preserving interval-valued moderate deviation function w.r.t. $\leq_{\alpha,\beta}$ for any $\alpha,\beta$.
  \item[(ii)] A class of $w$-preserving interval-valued moderate deviation functions w.r.t. $\leq_{\alpha,\beta}$ for any $\alpha,\beta$ can be obtained considering $\mathcal{D}_{\varepsilon,\delta}:[0,1]^2 \to \lr$ defined for positive constants $\varepsilon,\delta$ by (see Example 3.3 in \cite{decky2017}):
       \begin{equation}
       \mathcal{D}_{\varepsilon,\delta}(x,y)=\left\{
                                               \begin{array}{ll}
                                                 y-x+\varepsilon, & \textrm{ if } y>x, \\
                                                 0, &  \textrm{ if } y=x, \\
                                                 y-x-\delta, &  \textrm{ if } y<x
                                               \end{array}
                                             \right.
       \end{equation}
       and $C:[0,1]^2 \to [0,1]$ defined by $C(x,y)=\max\left(0,\min(1,f(y)-f(x)+y)\right)$, where $f:[0,1]\to \mathbb{R}$ is any non-decreasing function.

       Note that item (i) is a special case of item (ii) for $\varepsilon=\delta=0$ and $f=id$.
\end{enumerate}
\end{example}

\begin{example}\label{ex:conivmdf1}
In \cite{mymdv} (Theorem 6) a construction of a moderate deviation function $\mathcal{D}:[0,1]^2\to[-M_n,M_p]$ was introduced in the following way:
               \begin{equation}\label{eq:theqmdfref}
               \mathcal{D}(x,y)=
               \left\{
               \begin{array}{ll}
               M_p-M_pR_1(x,y), & \textrm{ if } x\leq y,\\[0.1cm]
               M_nR_2(x,y)-M_n, & \textrm{ if } x>y,
               \end{array}
               \right.
               \end{equation}
               for all $x,y\in[0,1]$, where $M_n,M_p\in]0,\infty[$. For different choices of restricted dissimilarity functions $R_1,R_2$ we obtain different moderate deviation functions. In particular, we give five examples, in each of them the choice of $M_p,M_n$ impact results where the ratio between $M_p$ and $M_n$ is important since it expresses the emphasis we put on the positive ($M_p$) or negative ($M_n$) deviation:

               \begin{enumerate}
               \item[(i)] If $R_1(x,y)=R_2(x,y)=1-|y-x|$, then
               \begin{equation}
               \mathcal{D}(x,y)=
               \left\{
               \begin{array}{ll}
               M_p(y-x), & \textrm{ if } x\leq y,\\[0.1cm]
               M_n(y-x), & \textrm{ if } x>y.
               \end{array}
               \right.
              \end{equation}

               \item[(ii)] If $R_1(x,y)=R_2(x,y)=1-|y^2-x^2|$, then
				\begin{equation}
               \mathcal{D}(x,y)=
               \left\{
                 \begin{array}{ll}
                  M_p(y^2-x^2), & \textrm{ if } x\leq y,\\[0.1cm]
                  M_n(y^2-x^2), & \textrm{ if } x>y.
                 \end{array}
               \right.
				\end{equation}

               \item[(iii)] If $R_1(x,y)=R_2(x,y)=1-(y-x)^2$, then
            \begin{equation}
               \mathcal{D}(x,y)=
               \left\{
                 \begin{array}{ll}
                  M_p(y-x)^2, & \textrm{ if } x\leq y,\\[0.1cm]
                  -M_n(y-x)^2, & \textrm{ if } x>y.
                 \end{array}
               \right.
               \end{equation}

               \item[(iv)] If $R_1(x,y)=1-|y^2-x^2|$ and $R_2(x,y)=1-(y-x)^2$, then
            \begin{equation}
               \mathcal{D}(x,y)=
               \left\{
                 \begin{array}{ll}
                  M_p(y^2-x^2), & \textrm{ if } x\leq y,\\[0.1cm]
                  -M_n(y-x)^2, & \textrm{ if } x>y.
                 \end{array}
               \right.
               \end{equation}

               \item[(v)] If $R_1(x,y)=1-(y-x)^2$ and $R_2(x,y)=1-|y^2-x^2|$, then
               \begin{equation}
               \mathcal{D}(x,y)=
               \left\{
                 \begin{array}{ll}
                  M_p(y-x)^2, & \textrm{ if } x\leq y,\\[0.1cm]
                  M_n(y^2-x^2), & \textrm{ if } x>y.
                 \end{array}
               \right.
               \end{equation}
               \end{enumerate}

Based on the approach given in Theorem~\ref{th_con1_iwmd}, we can build a $w$-preserving \noteb{interval-valued} moderate deviation function $D(L([0,1]))^2\to L(\mathbb{R})$ in such a way that we combine one of the five restricted dissimilarity functions $\mathcal{D}$ from items (i)-(v) with a function $C:[0,1]^2 \to [0,1]$ defined by
            \begin{equation}
            C(x,y)=\max\left(0,\min(1,f(y)-f(x)+y)\right)
            \end{equation}
            for some $f:[0,1]\to \mathbb{R}$ being a non-decreasing function (for example $Id$,\ldots).
\end{example}

It is worth to point out that $wD$-mean is based on the idea to use moderate deviation functions in a similar way as penalty functions are used to measure the similarity or dissimilarity between a given set of data~\cite{Calvo2010,Yager1993}. The main idea is, given a set of intervals, to determine another interval which represents all of them and which is the most similar to all of them in the sense determined by the moderate deviation function. That is, we are looking for the interval $Y$ which makes the sum $D(X_1,Y)+\ldots+D(X_n,Y)$ to be as close to $[0,0]$ as possible.

In what follows, we use our construction given by Theorem~\ref{th_con1_iwmd} to avoid the computation of $\sup$ and $\inf$ while obtaining $wD$-mean.

\begin{theorem}\label{th:mojemot}
Let $\alpha\in[0,1]$, $n\in\mathbb{N}$, let $M_p,M_n$ be positive real numbers and $\mathcal{D}:[0,1]^2\to[-M_n,M_p]$ be a moderate deviation function. Let $F:\li^{n+1}\to \mathbb{R}$ be the function given, for all $X_1,\ldots,X_n,Y\in\li$ such that $w(Y)=\min(w(X_1),\ldots,w(X_n))$, by:
\begin{equation}
F(X_1,\ldots,X_n,Y) = \mathcal{D}\Big(K_{\alpha}(X_1),K_{\alpha}(Y)\Big) + \ldots + \mathcal{D}\Big(K_{\alpha}(X_n),K_{\alpha}(Y)\Big).
\end{equation}
Then
\begin{enumerate}
  \item[(i)] If $\mathcal{D}$ is continuous, then, for each $n$-tuple $(X_1,\ldots,X_n)\in\li^n$, there exists $Y\in\li$ such that $w(Y)=\min(w(X_1),\ldots,w(X_n))$ and $F(X_1,\ldots,X_n,Y)=0$.
  \item[(ii)] If $\mathcal{D}$ is strictly increasing in the second component, then, for each $n$-tuple $(X_1,\ldots,X_n)\in\li^n$, there exists at most one $Y\in\li$ such that $w(Y)=\min(w(X_1),\ldots,w(X_n))$ and $F(X_1,\ldots,X_n,Y)=0$.
\end{enumerate}
\end{theorem}
\dem
(i) Since the continuity of $\mathcal{D}$ implies the continuity of the function $F(X_1,\ldots,X_n,\cdot)$, the proof follows from the observation that $F(X_1,\ldots,X_n,Z_1)\leq 0$ and $F(X_1,\ldots,X_n,Z_2)\geq 0$ if $K_{\alpha}(Z_1)=\min\left(K_{\alpha}(X_1),\ldots,K_{\alpha}(X_n)\right)$ and $K_{\alpha}(Z_2)=\max\left(K_{\alpha}(X_1),\ldots,K_{\alpha}(X_n)\right)$. Note that since $w(Y)$ is fully determined by fixed $n$-tuple $(X_1,\ldots,X_n)$, the continuity of the function $F(X_1,\ldots,X_n,\cdot)$ is considered in the sense of the standard continuity of a real function with the variable $K_{\alpha}(Y)$.

(ii) Observe that the strict monotonicity of $D$ implies the strict monotonicity (increasingness) of the function $F(X_1,\ldots,X_n,\cdot)$. Again in the sense of a real function with the variable $K_{\alpha}(Y)$.
\deme

\begin{comment}
The previous theorem could be formulated in a more general way: instead of $w(Y)=\min(w(X_1),\ldots,w(X_n))$ there can be $w(Y)=B(w(X_1),\ldots,w(X_n))$ for any aggregation function $B$, but then the domain of $Y$ should be extended, in particular, $Y\in L(\mathbb{R})$ in the case of any $B$ and not $Y\in\li$ as it is in the case of minimum, although $K_{\alpha}(Y)\in[0,1]$ in both cases. Since our goal is to show that $Y$ is an output of $wD$-mean, we used this version of the theorem to keep the range of $wD$-mean in $\li$.
\end{comment}

The following corollary gives us a method of constructing $wD$-means based on Theorem~\ref{th_con1_iwmd}, Theorem~\ref{th:mojemot} and Example~\ref{ex:conivmdf1}.

\begin{corollary}\label{cor:mojemot}
Under the assumptions of Theorem~\ref{th:mojemot}, where $\mathcal{D}:[0,1]^2\to \mathbb{R}$ is given by Equation~\eqref{eq:theqmdfref} with $R_1,R_2$ being continuous strictly monotone restricted equivalence functions and $D:(\li)^2\to L(\mathbb{R})$ is given by Theorem~\ref{th_con1_iwmd}, the following statements are equivalent:
\begin{enumerate}
  \item[(i)] $F(X_1,\ldots,X_n,Y)=0$;
  \item[(ii)] $M_D(X_1,\ldots,X_n)=Y$, where the \noteb{interval-valued} $wD$-mean $M_D$ is given by Equation~\eqref{eq_wmd} with $B=\min$;
  \item[(iii)] \begin{multline}\label{eq:eqivwdmean}
                \sum\limits_{i=1}^{k}\bigg(M_p-M_pR_1\Big(K_{\alpha}\left(X_{\sigma(i)}\right),K_{\alpha}\left(Y\right)\Big)\bigg) \\+ \sum\limits_{i=k+1}^{n}\bigg(M_nR_2\Big(K_{\alpha}\left(X_{\sigma(i)}\right),K_{\alpha}\left(Y\right)\Big)-M_n\bigg) = 0
            \end{multline}
\end{enumerate}
where $\sigma:\{1,\ldots,n\}\to\{1,\ldots,n\}$ is a permutation such that $X_{\sigma(1)}\leq_{\alpha,\beta}\ldots\leq_{\alpha,\beta} X_{\sigma(n)}$ and $k$ is the greatest number from $\{1,\ldots,n\}$ satisfying
\begin{equation}\label{eq:lkmnas}
\sum\limits_{i=1}^{n} \mathcal{D}(K_{\alpha}(X_{\sigma(i)}),K_{\alpha}(X_{\sigma(k)}))\leq 0.
\end{equation}
Moreover, $K_{\alpha}(Y)\in[K_{\alpha}(X_{\sigma(k)}),K_{\alpha}(X_{\sigma(k+1)})[$ whenever $k<n$ and $K_{\alpha}(Y)=K_{\alpha}(X_{\sigma(1)})=\ldots=K_{\alpha}(X_{\sigma(n)})$ whenever $k=n$.
\end{corollary}
\dem
First observe that, due to $(MD1)$ and $(MD2)$, we have: if $k$ satisfies Equation~\eqref{eq:lkmnas}, then for all $p\in\{1,\ldots,k\}$ and all $q\in\{k+1,\ldots,n\}$ it holds
\begin{multline}
\sum\limits_{i=1}^{n} \mathcal{D}(K_{\alpha}(X_{\sigma(i)}),K_{\alpha}(X_{\sigma(p)})) \leq 0 \qquad\\ \textrm{ and } \qquad \sum\limits_{i=1}^{n} \mathcal{D}(K_{\alpha}(X_{\sigma(i)}),K_{\alpha}(X_{\sigma(q)})) > 0.
\end{multline}
Then the equivalence of (i) and (ii) follows from Theorem~\ref{th:mojemot} and the equivalence of (i) and (iii) follows from the observation:
\begin{multline}
F(X_1,\ldots,X_n,Y) = \sum\limits_{i=1}^{k}\bigg(M_p-M_pR_1\Big(K_{\alpha}\left(X_{\sigma(i)}\right),K_{\alpha}\left(Y\right)\Big)\bigg) +\\ \sum\limits_{i=k+1}^{n}\bigg(M_nR_2\Big(K_{\alpha}\left(X_{\sigma(i)}\right),K_{\alpha}\left(Y\right)\Big)-M_n\bigg).
\end{multline}
\deme

\begin{example}\label{ex:conivmdf3}
For each particular choice of $\mathcal{D}$ (or $R_1,R_2$) in cases (i)-(v) of Example~\ref{ex:conivmdf1}, the Equation~\eqref{eq:eqivwdmean} has the following form:
\begin{enumerate}
\item[(i)]
\begin{multline*}
\sum\limits_{i=1}^{k}\bigg(M_p\Big(K_{\alpha}\left(Y\right)-K_{\alpha}\left(X_{\sigma(i)}\right)\Big)\bigg) + \\ \sum\limits_{i=k+1}^{n}\bigg(M_n\Big(K_{\alpha}\left(Y\right)-K_{\alpha}\left(X_{\sigma(i)}\right)\Big)\bigg) = 0
\end{multline*}
and the solution is:
$$
K_{\alpha}(Y) = \frac{M_p\sum\limits_{i=1}^{k}K_{\alpha}\left(X_{\sigma(i)}\right)+M_n\sum\limits_{i=k+1}^{n}K_{\alpha}\left(X_{\sigma(i)}\right)}{kM_p+(n-k)M_n}
$$

\item[(ii)]
\begin{multline*}
\Big(K_{\alpha}\left(Y\right)\Big)^2\Big(kM_p+(n-k)M_n\Big) - \\ M_p\sum\limits_{i=1}^{k}\Big(K_{\alpha}\left(X_{\sigma(i)}\right)\Big)^2 -  M_n\sum\limits_{i=k+1}^{n}\Big(K_{\alpha}\left(X_{\sigma(i)}\right)\Big)^2 = 0
\end{multline*}
and the solution is:
$$
K_{\alpha}\left(Y\right) = \sqrt{\frac{M_p\sum\limits_{i=1}^{k}\Big(K_{\alpha}\left(X_{\sigma(i)}\right)\Big)^2 + M_n\sum\limits_{i=k+1}^{n}\Big(K_{\alpha}\left(X_{\sigma(i)}\right)\Big)^2}{kM_p+(n-k)M_n}}
$$

\item[(iii)]
\begin{multline*}
	\Big(K_{\alpha}\left(Y\right)\Big)^2\Big(kM_p-(n-k)M_n\Big) + \\ K_{\alpha}\left(Y\right)\Big(2M_n\sum\limits_{i=k+1}^{n}K_{\alpha}\left(X_{\sigma(i)}\right)-2M_p\sum\limits_{i=1}^{k}K_{\alpha}\left(X_{\sigma(i)}\right)\Big) +
	\\
	M_p\sum\limits_{i=1}^{k}\Big(K_{\alpha}\left(X_{\sigma(i)}\right)\Big)^2 - M_n\sum\limits_{i=k+1}^{n}\Big(K_{\alpha}\left(X_{\sigma(i)}\right)\Big)^2 = 0
\end{multline*}

\item[(iv)]
\begin{multline*}
\Big(K_{\alpha}\left(Y\right)\Big)^2\Big(kM_p-(n-k)M_n\Big) + \\ K_{\alpha}\left(Y\right)\Big(2M_n\sum\limits_{i=k+1}^{n}K_{\alpha}\left(X_{\sigma(i)}\right)\Big) +
\\
M_p\sum\limits_{i=1}^{k}\Big(K_{\alpha}\left(X_{\sigma(i)}\right)\Big)^2 - M_n\sum\limits_{i=k+1}^{n}\Big(K_{\alpha}\left(X_{\sigma(i)}\right)\Big)^2 = 0
\end{multline*}

\item[(v)]
\begin{multline*}
\Big(K_{\alpha}\left(Y\right)\Big)^2\Big(kM_p-(n-k)M_n\Big) + \\ K_{\alpha}\left(Y\right)\Big(-2M_p\sum\limits_{i=1}^{k}K_{\alpha}\left(X_{\sigma(i)}\right)\Big) +
\\
M_p\sum\limits_{i=1}^{k}\Big(K_{\alpha}\left(X_{\sigma(i)}\right)\Big)^2 - M_n\sum\limits_{i=k+1}^{n}\Big(K_{\alpha}\left(X_{\sigma(i)}\right)\Big)^2 = 0
\end{multline*}
\end{enumerate}
\end{example}

\section{\notea{Interval-valued aggregation functions and BCI frameworks}} \label{sec:md_bci}
\notea{In this section we present the two MI BCI frameworks we have used in our experimentation: the Traditional Framework in Section \ref{sec:tradBCI} and the Multimodal Fuzzy Fusion framework in Section \ref{sec:mff}. Then, we explain how we apply the Interval-valued moderate deviations in both cases, in Section \ref{sec:md_applied}.}
\subsection{Traditional BCI framework}
\label{sec:tradBCI}

\begin{figure*}
	\includegraphics[width=\linewidth]{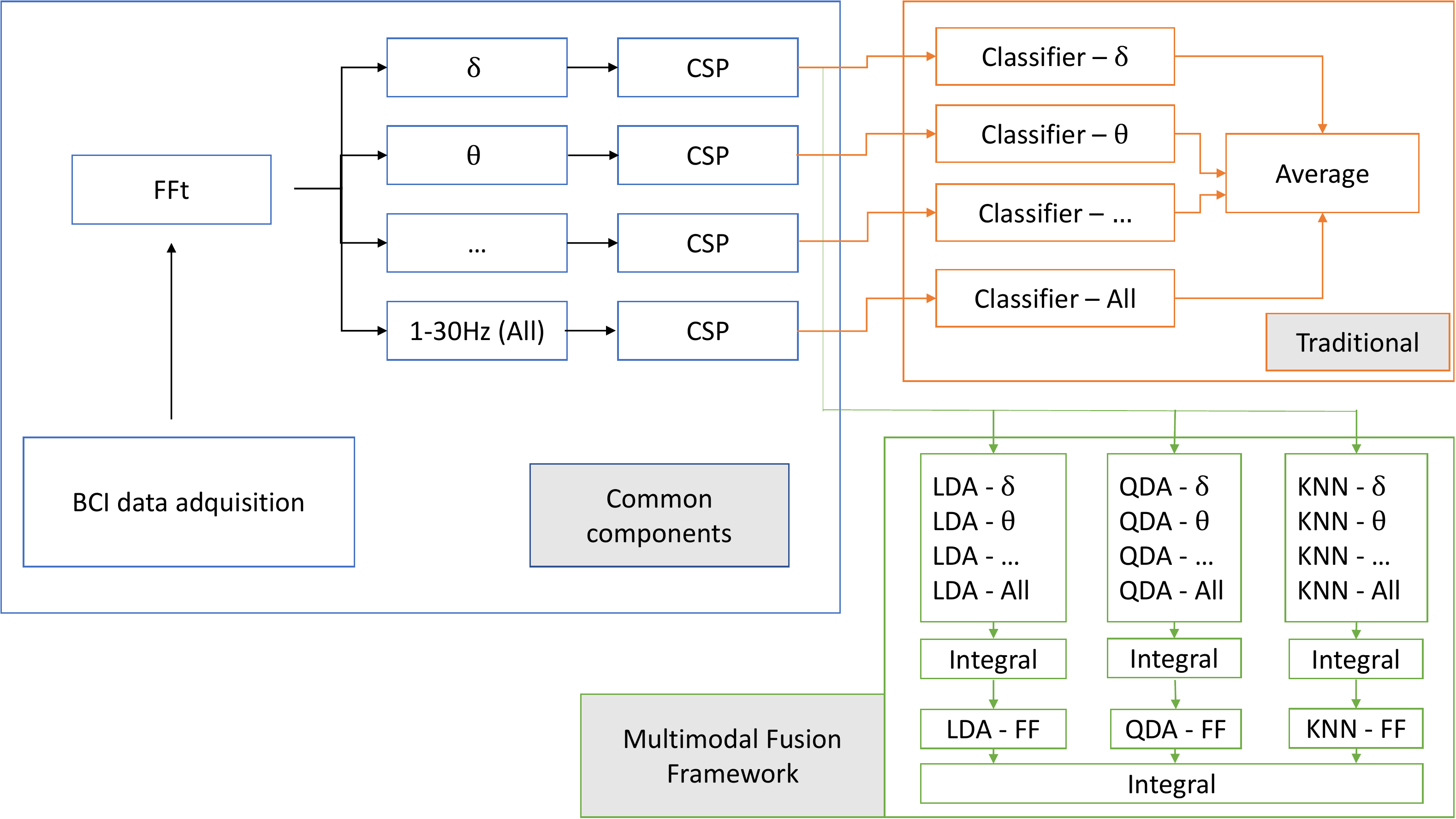}
	\caption{Components of the Traditional and Multimodal Fusion BCI Frameworks.}
	\label{fig:frameworks}
\end{figure*}
The traditional BCI system was proposed in \cite{wu2016fuzzy}. Its structure includes four parts (Fig. \ref{fig:frameworks}):

\begin{enumerate}
	\item The first step is acquiring the EEG data from a EEG device and performing band-pass filtering	and artefact removal on the collected EEG signals.
	
	\item The second
	step is EEG feature transformation and feature extraction. Usually, the FFt is used to rapidly transform
	the EEG signals into different frequency components \cite{akin2002comparison}. The FFt
	analysis transforms the time-series EEG signals in each channel into the specified frequency range, which in our case is from 1 to 30 Hz, covering the delta ($\delta$) 1-3 Hz, theta ($\theta$) 4-7 Hz, alpha ($\alpha$) 8-13 Hz, beta ($\beta$) 14-30 Hz and \notea{\textit{All}} 1-30Hz bands using a 50-point moving window segment overlapping 25 data points. \notea{Although some redundancy is included in the system, the \textit{All} band is considered in order to study possible interactions among non-adjacent frequencies and to gather additional features for the subsequent CSP and classifiers.}
	
	\item Subsequently, the CSP is
	used for feature extraction to extract the maximum spatial separability from the different EEG signals corresponding to the control commands. The CSP is a well-known supervised mathematical procedure commonly used in EEG signal processing. The CSP is used to transform multivariate EEG signals into well-separated subcomponents with maximum spatial variation using the labels for each example \cite{blankertz2007optimizing}, \cite{guger2000real}, \cite{gramfort2013meg}.
	
	\item Last, pattern classification is performed on the
	CSP features signals using an ensemble of classifiers
	to differentiate the commands. Each base classifier is trained using a different wave band (for instance, if the base classifier is the LDA, the ensemble would be composed of: $\delta-LDA$, $\theta-LDA$, $\alpha-LDA$, $\beta-LDA$, and $All-LDA$) and the final decision is taken combining all of them. The most common way of obtaining the final decision is to compute the arithmetic mean of the outputs of all the base classifiers (each one provides a probability for each class), and take the class with highest aggregated value. Some of the most common classifiers used for this task are the LDA,
	QDA and KNN \cite{tibarewala2010performance}. This part would correspond to the orange box in Fig. \ref{fig:frameworks}, labeled as ``Traditional''.
\end{enumerate}

For this work, we have used \notea{LDA} classifiers in the traditional framework, as they are very popular in the BCI literature \cite{wu2013common, kee2012genetic, tsui2008comparison, naseer2016determining, geng2008novel}.

\subsection{Multimodal Fuzzy Fusion BCI framework} \label{sec:mff}
The Multi-modal Fuzzy Framework (MFF) is proposed in \cite{ko2019multimodal}. It follows a similar structure to the one in the traditional BCI framework (Fig. \ref{fig:frameworks}): it starts with the EEG measurements, it computes the FFt transformation to the frequency domain and it uses the CSP transform to obtain a set of features to train the classifiers.

However, in the MFF it is necessary to train not one, but three classifiers for each wave band: a LDA, a QDA and a KNN. We name the classifiers according to their type of classifier and the wave band used to train it. For instance, for the $\delta$ band we would have $\delta-LDA$, $\delta-QDA$ and $\delta-KNN$.

Then, the decision making phase is performed in two phases (This part would correspond to the green box in Fig. \ref{fig:frameworks}, labeled as ``Multimodal Fusion Framework''):
\begin{enumerate}
	\item Frequency phase: since we got a LDA, QDA and KNN for each wave band, the first step is to fuse the outputs of these classifiers in each wave band. For example, in the case of the LDA classifiers, we have a $\delta-LDA$, $\theta-LDA$, $\alpha-LDA$, $\beta-LDA$ and $All-LDA$ that will be fused them using an aggregation function to obtain a vector, $FF-LDA$. That is, the same process explained for the traditional framework is applied but without making the final decision. We do the same with the QDA and KNN classifiers. The result of this phase is a list of collective vectors (one for each type of classifier).
	\item Classifier phase: in this phase, the input is the list of collective vectors given by each different kind of classifier ($FF-LDA$, $FF-LDA$, $FF-KNN$) computed in the frequency phase. We fuse the three vectors according to the classes, and the result is a vector containing the score for each class for the given sample. As in the traditional framework, the decision is made in favour to the class associated with the largest value.
\end{enumerate}
We must point out that the same aggregation is used for both the frequency phase and the classifier phases.

The aggregation functions tested in the MFF are the Choquet integral, the CF integral using the Hamacher T-norm, the $CF_{min,min}$ generalizations, the Sugeno integral and the Hamacher Sugeno integral \cite{ko2019multimodal, lucca2018cf, dimuro2020state}.

\subsection{\notea{Interval-valued Moderate Deviations applied in the BCI frameworks}} \label{sec:md_applied}

We have tested the interval-valued moderate deviation functions in the two different MI BCI frameworks previously introduced. The idea in both cases is to replace the existing aggregations for the classifier outputs (the arithmetic mean in the traditional and the fuzzy integrals in the MFF) for our new developed ones.

First, we construct the intervals from the probability for each class obtained from each classifier. We use the length of the intervals to measure the inaccuracies or uncertainties related to these classifiers' outputs. To do so, we have used the mapping in \cite{Jurio2011}:
\begin{equation}
	F(x,y) = [\underline{I(x,y)}, \overline{I(x, y)+y}]
\end{equation}
where I is an \notea{fuzzy implication function}, $x$ is the probability for each class obtained from the classifier and we set $y$ as $0.3$. We crop the values so that they are contained in the $[0,1]$ interval. We have tried three different \notea{fuzzy implication function}s to construct the  intervals: the \L ukasiewicz \notea{fuzzy implication function}, the Reichenbach \notea{fuzzy implication function} and the Kleene-Dienes \notea{fuzzy implication function}.

Then, we aggregate the interval-valued logits from the classifiers using an interval-valued moderate deviation based function. We have constructed the deviation function using the Eq. (\ref{eq:theqmdfref}). We have named $MD_1$ to the interval-valued moderate deviation using $R_1(x,y)=R_2(x,y)=1-\left|y-x\right|$ and the $MD_2$ setting $R_1=1-(y-x)^2$ and $R_2(x,y)=1-|y^2-x^2|$.

In the traditional BCI framework, we follow this interval construction algorithm and the $MD_1$ and $MD_2$ respectively to substitute the ``Average" block in Figure \ref{fig:frameworks}. In the case of the MFF we use them to substitute all the ``Integral" blocks in Fig. \ref{fig:frameworks}.

To construct the interval-valued moderate deviation functions we need to set the parameters $M_p$ and $M_n$. We have optimized them by taking 200 samples of different pairs in the $[1,100]$ range and testing the accuracy against the training set. We have opted for this method as it seemed to surpass other optimization algorithms in terms of computational time, obtaining similar accuracy results.

\section{Experimental Results in Motor-Imagery based BCI} \label{sec:md_experiments}
In this section we discuss the behaviour of our new approaches in the BCI competition IV dataset 2a \noteb{(IV-2a)} \noteb{and the BCI competition IV dataset 2b (IV-2b), which are detailed in \cite{brunner2007spatial}}. The IV-2a dataset consists of four classes of MI tasks: tongue, foot, left-hand and right-hand performed by 9 volunteers. For each task, 22 EEG channels were collected. There is a total of 288 trials for each participants, equally distributed among the 4 classes. \noteb{The IV-2b dataset consists of three different EEG channels for each subject, who performed two different motor imagery tasks: moving the right hand or moving the left hand.}

For our experimental setup, we have used 4 out of the 22 channels available in the IV-2a dataset, \noteb{(C3, C4, CP3, CP4)}, as they reported good results in \cite{ko2019multimodal}. \noteb{For the IV-2b dataset, we have used the three channels available}. As features, we have used the FFt to obtain the $\delta$, $\theta$, $\alpha$, $\beta$, \noteb{SensoriMotor Rhythm (SMR) ($13-15$Hz)} and All, and we have used a CSP filter with 25 components. From each subject, we have generated twenty partitions (50\% train and 50\% test). So, this produces a total of 180 datasets. The accuracy for each framework is computed as the average for all of them.

\noteb{For the IV-2a,} we have studied the full four classes task, and we have also studied the binary classification task Left hand vs Right hand, as it is common practice in the literature \cite{ko2019multimodal, jafarifarmand2017new, aghaei2015separable}. \notea{Results for each individual subject are available at \url{https://github.com/Fuminides/interval_md_bci_results}}.

\subsection{Left/Right hand task} \label{subsec:binary}
In Table \ref{tab:md2_binary} we have shown the results obtained using the $MD_1$ aggregation in the Left/Right hand task. We have used \notea{the three considered} \notea{fuzzy implication function}s and the two possible MI BCI frameworks. We found that the \L ukasiewicz \notea{fuzzy implication function} is the one that works best for both frameworks \noteb{in the IV-2a dataset}. \noteb{For the IV-b dataset, we found the Reichenbach operator is the best one in the traditional framework and the three operators gave very similar results for the MFF.}

In Table \ref{tab:md5_binary} we have displayed the results using the $MD_2$ aggregation function. We found that in this case that \noteb{in the IV-a dataset} the best \notea{fuzzy implication function} is the Reichenbach \notea{fuzzy implication function} in the traditional framework and the \L ukasiewicz one for the MFF. Results are higher or equal in all cases using this function compared to the MD$_1$. The best result is the 0.8251 obtained using the MD$_2$ traditional framework with the Reichenbach \notea{fuzzy implication function}. \noteb{In the IV-2b dataset, we found very similar performance for any of the implication operators and we also noted that the MFF performed better in all cases than the traditional framework.}

%The best results for this task, $0.8251$, were achieved using the traditional framework, the Reichenbach \notea{fuzzy implication function} and the MD$_2$.
%In both Table \ref{tab:md2_binary} and \ref{tab:md5_binary} we observe that the traditional framework is more sensible to the \notea{fuzzy implication function} used than the MFF framework. In both the case of the $MD_1$ and the $MD_2$ the lowest performance of the MFF is better than the lowest performance of the traditional framework, but the best result is obtained with the traditional framework.

\begin{table}
  \centering
  \caption{\noteb{Accuracies for the binary task using the $MD_1$ as aggregation function.}}
  \begin{tabular}{lcc}
	\toprule
	{} &     Trad. &       MFF \\
	\midrule
	Kleene-Dienes &  $0.7443\pm0.1840$ &  $0.7513\pm0.1906$\\
	Lukasiewicz   &  $0.7548\pm0.1962$ &  $0.7566\pm0.1871$\\
	Reichenbach   &  $0.7443\pm0.1840$ & $0.7513\pm0.1906$ \\
	\bottomrule
\end{tabular}

  \label{tab:md2_binary}
\end{table}
\begin{table}
  \centering
  \caption{\noteb{Accuracies for the binary task using the $MD_2$ as aggregation function.}}
  \begin{tabular}{lcc}
	\toprule
	{} &     Trad. &       MFF \\
	\midrule
	Kleene-Dienes &  $0.7443\pm0.1840$ & $0.7513\pm0.1906$ \\
	Lukasiewicz   &  $0.8146\pm0.0541$ & $0.8191\pm0.0395$ \\
	Reichenbach   &  $0.8251\pm0.0538$ & $0.8184\pm0.0415$ \\
	\bottomrule
\end{tabular}
  \label{tab:md5_binary}
\end{table}

\subsection{Four classes task} \label{subsec:four}
In Table \ref{tab:md2_full} we have shown the results obtained using the $MD_1$ aggregation in the left hand, right hand, tongue and foot task for \notea{the three considered fuzzy implication function}s and the two frameworks. In this case we obtained the best result using the traditional framework and the \L ukasiewicz \notea{fuzzy implication function}.

In Table \ref{tab:md5_full} we have shown the results obtained using the $MD_2$ aggregation for all the \notea{fuzzy implication function}s and the two frameworks. The MFF framework achieved its best result using the \L ukasiewicz \notea{fuzzy implication function}. The MD$_2$ always performed better the the MD$_1$ in this case too.

The best results for this task, $0.6943$, were achieved using the MFF, the \L ukasiewicz and the MD$_2$.

\begin{table}
  \centering
  \caption{\noteb{Accuracies for the four classes task using the $MD_1$ as aggregation function.}}
  \begin{tabular}{lcc}
\toprule
{} &     Trad. &       MFF \\
\midrule
Kleene-Dienes & $0.6256\pm0.0352$ & $0.6217\pm0.1904$ \\
Lukasiewicz   &  $0.6358\pm0.0366$ & $0.6073\pm0.2112$ \\
Reichenbach   &  $0.6256\pm0.0352$ & $0.6217\pm0.1904$ \\
\bottomrule
\end{tabular}

  \label{tab:md2_full}
\end{table}

\begin{table}
  \centering
  \caption{\noteb{Accuracies for the four classes task using the $MD_2$ as aggregation function.}}
  \begin{tabular}{lcc}
\toprule
{} &     Trad. &       MFF \\
\midrule
Kleene-Dienes &  $0.6256\pm0.0352$ & $0.6217\pm0.1904$ \\
Lukasiewicz   &  $0.6406\pm0.0396$ & $0.6943\pm0.0338$ \\
Reichenbach   &  $0.6471\pm0.0359$ & $0.6905\pm0.0328$ \\
\bottomrule
\end{tabular}

  \label{tab:md5_full}
\end{table}
\subsection{\notea{Comparison to the non-interval-valued case}}
We have compared the best interval-valued MD aggregation, the MD$_2$, with the standard arithmetic average aggregation in the traditional framework and the Choquet integral in the MFF framework, as it was the best aggregation for the MFF in \cite{ko2019multimodal}.

In Table \ref{tab:binary_numeric_performances} there are the results for the Traditional and MFF frameworks in the Left/Right hand task without using interval-valued aggregations, and the best result found in Sections \ref{subsec:binary} and \ref{subsec:four}. In this case, we obtained a moderate increase in accuracy compared to the numerical results \noteb{for the case of IV-2a dataset, but not in the case of the IV-b, in which the Choquet integral performed best}. We also performed an analogous study for the four classes task, reported in Table \ref{tab:full_task_numeric_performances}. We found in this case that all the configurations performed almost equally, although the MD$_2$ performed slightly better than the Choquet MFF.

In Tab. \ref{tab:p_value_numerical_binary}, we have computed the statistical differences \notea{for the left/right hand task} among the numerical aggregations and the MD$_2$. We have found significant differences favoring the MD$_2$ aggregation \noteb{for the case of the IV-2a dataset and the Choquet MFF for the IV-2b dataset} \notea{using Kruskal test and the Anderson post-hoc to compute the respective $P$-values}. We have performed the analogous study for the four classes task in Tab. \ref{tab:p_value_numerical_four_classes}. We have found in that case that the difference favouring the MD$_2$ is not significant compared to the Choquet integral in the MFF, but it is in the case of the arithmetic mean in the traditional framework.

\begin{table}
	\centering
	\caption{\noteb{Performance for the Left/Right classes task using different aggregation functions.}}
	\begin{tabular}{llc}
		\toprule
		\noteb{Dataset} & Classifier &           Accuracy \\
		\midrule
	& 	Trad. Average &  $0.8119\pm0.0384$ \\
	\noteb{IV-2a}& 	MFF Choquet & $0.8049\pm0.0417$\\
	&	MD$_2$ (Trad.) &  $0.8251\pm0.0538$ \\
	\midrule
		& 	Trad. Average &  $0.7213\pm 0.0824$ \\
	\noteb{IV-2b}& 	MFF Choquet & $0.7440\pm 0.0058$\\
	&	MD$_2$ (Trad.) &  $0.7366\pm0.0708 $ \\
		\bottomrule
	\end{tabular}
	\label{tab:binary_numeric_performances}
\end{table}

%\begin{table}
%	\centering
%	\begin{tabular}{lc}
%		\toprule
%		Classifier &           Accuracy \\
%		\midrule
%		MD$_1$ (MFF.) &  $0.7566\pm0.1871$ \\
%		MD$_2$ (Trad.) &  $0.8251\pm0.0538$ \\
%		\bottomrule
%	\end{tabular}
%	\caption{Performance for the Left/Right hand task using the traditional framework.}
%	\label{tab:binary_intervalar_performances}
%\end{table}

\begin{table}
	\centering
	\caption{\noteb{Performance for the four classes task using different aggregation functions.}}
	\begin{tabular}{llc}
		\toprule
		\noteb{Dataset} & Classifier &           Accuracy \\
		\midrule
		 & Average (Trad) &  $0.6494\pm0.0369$ \\
	\noteb{IV-2a} & 	Choquet (MFF) & $0.6919\pm0.0314$\\
	&	MD$_2$ (MFF) &  $0.6943\pm0.0338$ \\
		\bottomrule
	\end{tabular}
	
	\label{tab:full_task_numeric_performances}
\end{table}

\begin{table}
	\centering
	\caption{\noteb{$P$-values for the Left/Right hand task, comparing numerical aggregations to the MD$_2$. $*$ marks statistical differences ($P$-value lesser than 0.05).}}
	\begin{tabular}{llll}
		\toprule
		 \noteb{Dataset} & {} & Trad. Average & MFF Choquet \\
		\midrule
		%Trad. Average &               &             &      \\
		\noteb{IV-2a} & MFF Choquet   &         $.07$ &                  \\
		& Md$_2$          &           $*$ &         $*$      \\
		\midrule
		 & {} & Trad. Average & MFF Choquet \\
		\midrule
		%Trad. Average &               &             &      \\
		\noteb{IV-2b} & MFF Choquet   &         $*$ &                  \\
		& Md$_2$          &           $*$ &         $*$      \\
		
		\bottomrule
	\end{tabular}
	
	\label{tab:p_value_numerical_binary}
\end{table}

\begin{table}
	\centering
	\caption{\noteb{$P$-values for the four classes task, comparing numerical aggregations to the MD$_2$. $*$ marks statistical differences ($P$-value lesser than 0.05).}}
	\begin{tabular}{llll}
		\toprule
		\noteb{Dataset} & {} & Trad. Average & MFF Choquet \\
		\midrule
		\noteb{IV-2a} & MFF Choquet   &           $*$ &                   \\
	&	Md$_2$          &         $*$ &         $.25$       \\
		\bottomrule
	\end{tabular}

	\label{tab:p_value_numerical_four_classes}
\end{table}
%\begin{table}
%	\centering
%	\begin{tabular}{lc}
%		\toprule
%		Classifier &           Accuracy \\
%		\midrule
%		MD$_1$ (MFF) &  $0.6217\pm0.1904$ \\
%		MD$_2$ (MFF) &  $0.6905\pm0.0328$ \\
%		%OWA1 &  $0.6686\pm0.0353$ \\
%		%OWA2 &  $0.6731\pm0.0344$ \\
%		%OWA3 &  $0.6494\pm0.0374$ \\
%		\bottomrule
%	\end{tabular}
%	\caption{Performance for the four classes task using the interval-valued moderate deviation based aggregations.}
%	\label{tab:md_performances}
%\end{table}
\subsection{Comparison against interval-valued OWA operators}

To compare the results obtained using the interval-valued moderate deviations with other interval-valued aggregations, we have compared them against the results obtained using interval-valued OWA operators. We have used three OWA operators: OWA$_1$, OWA$_2$ and OWA$_3$, as described in Section \ref{sec:owa}. We have computed the Kruskal statistical test and the Anderson post-hoc to look for significant differences among each method, using a level of significance of 0.05.

In Tab. \ref{tab:left_right_comparison} we show the comparison among the interval-valued OWAs and the MD$_2$ for the left/right hand task. We show for each aggregation the result for the best configuration (\notea{each configuration is composed of one the three different fuzzy implication functions and one of the two different frameworks}) for each aggregation. We found that the $MD_2$ is the best aggregation for this task, followed by $OWA_2$, $OWA_1$, $OWA_3$. The Kruskal test found statistical differences among the different aggregations, and the results for the Anderson post-hoc are in Tab. \ref{tab:binary_p-values}, which shows that the MD$_2$ performs statistically better than the rest of the aggregations.

In Tab. \ref{tab:four_classes_comparison} we display the comparison among the interval-valued OWAs and the MD$_2$ for the four classes classification task. The results do not change much when it comes to decide which aggregation is the best, and the MD$_2$, is again the best aggregation.
The Kruskal test found statistical differences among the different aggregations, and the results for the post-hoc are in Tab. \ref{tab:four_classes_p-values}, which shows that the MD$_2$ performs statistically better than the rest of the aggregations, just as in the Left/Right hand task.

%We have computed the Kruskal statistical test and it's respective Anderson post-hoc, to check significant statistical differences among the interval-valued OWA operators, in both the binary and four classes tasks. We have shown the post-hoc results in Tables \ref{tab:binary_p-values} and \ref{tab:four_classes_p-values}. We have found that the MD$_2$ performance advantage is statistically significant in all cases.

%We have computed the results for both the traditional and MFF BCI frameworks using non interval-valued methods. In the case of the MFF we have used the Choquet integral, as it was the best aggregation reported in \cite{ko2019multimodal}, and in the case of the traditional framework we have used the arithmetic mean. We found that the $MD_2$ configuration obtained here surpassed

%\begin{table}
%  \begin{tabular}{lc}
%    \toprule
%    Configuration & Accuracy \\
%    \midrule
%    Traditional framework using Arithmetic mean & 81.16 \\
%    MFF using Choquet Integral & 80.02 \\
%   MFF using $MD_2$ \& Reichenbach & 81.45 \\
%    \bottomrule
%  \end{tabular}
%  \caption{Comparison of the best interval-valued configuration obtained against the non interval-valued aggregation functions for the left/right hand task.}
%  \label{tab:comparison_std_results}
%\end{table}

\begin{table}
  \centering
  \caption{\noteb{Best accuracy for each different aggregation functions in the Left/Right hand task.}}
  \begin{tabular}{llc}
  	\toprule
	\noteb{Dataset} & 	Classifier &           Accuracy \\
  	\midrule
  	%MD$_1$ &  $0.7566\pm0.1871$ \\
  	\noteb{IV-2a} & MD$_2$ &  $0.8251\pm0.0415$ \\
&  	OWA$_1$ &  $0.8044\pm0.0476$ \\
&  	OWA$_2$ &  $0.8086\pm0.0465$ \\
&  	OWA$_3$ &  $0.7993\pm0.0483$ \\
\midrule
	\noteb{IV-2b} & MD$_2$ &  $0.7366 \pm 0.0708 $ \\
&  	OWA$_1$ &  $0.7300\pm0.0800$ \\
&  	OWA$_2$ &  $0.7312\pm0.0285$ \\
&  	OWA$_3$ &  $0.7296\pm0.0786$ \\

  	\bottomrule
  \end{tabular}
  
  \label{tab:left_right_comparison}
\end{table}

%\begin{figure}
%  \centering
%  %\includegraphics[width=\linewidth]{gallery/accuracies_binary_average.pdf}
%  \begin{tabular}{cc}
%  	\toprule
%  	Aggregation Function & Accuracy \\
%  	\midrule
%  	OWA$_1$ & 0.7590 \\
%  	OWA$_2$ & 0.7920 \\
%  	OWA$_3$ & 0.7997 \\
%  	MD$_1$ & 0.8012 \\
%  	MD$_2$ &  0.8022 \\
%  	\bottomrule
%  \end{tabular}
%  \caption{Average accuracy for each different configuration using different aggregation functions in the left/right hand task.}
%  \label{fig:left_right_comparison_average}
%\end{figure}

\begin{table}
	\centering
	
	\caption{\noteb{$P$-values for the Left/Right hand task comparison for the interval-valued aggregations. $*$ marks statistical differences ($P$-value lesser than 0.05).}}
	
	\begin{tabular}{llllll}
		\toprule
		\noteb{Dataset} & {Trad.} &  MD$_2$ & OWA$_1$ &   OWA$_2$ &   OWA$_2$ \\
		\midrule
		%MD$_1$  &  $*$ &  $*$ &    $*$ &    $*$ \\
		\noteb{IV-2a} &		MD$_2$  &      &  $*$ &    $*$ &    $*$ \\
		%\hdashline
		&		OWA$_1$ &      &      &  $.25$ &  $.25$ \\
		&		OWA$_2$ &      &      &        &  $.25$ \\
		%\bottomrule
	\end{tabular}
	
	\begin{tabular}{llllll}
		\toprule
		\phantom{Dataset}  & {MFF} &  MD$_2$ & OWA$_1$ &   OWA$_2$ &   OWA$_2$ \\
		\midrule
		%MD$_1$  &  $*$ &  $.22$ &  $.16$ &  $.09$ \\
		\phantom{IV-2a} &		MD$_2$  &      &    $*$ &  $*$ &    $*$ \\
		%\hdashline
		&		OWA$_1$ &      &        &  $.25$ &  $.25$ \\
		&		OWA$_2$ &      &        &        &  $.13$ \\
		\bottomrule
	\end{tabular}
	
	\label{tab:binary_p-values}
\end{table}

\begin{table}
  \centering
   \caption{\noteb{Best accuracy for each different aggregation functions in the four classes hand task.}}
  \begin{tabular}{llc}
  	\toprule
  	\noteb{Dataset} & Classifier &           Accuracy \\
  	\midrule
  	%MD$_1$ &  $0.6358\pm0.0366$ \\
  	\noteb{IV-2a} &MD$_2$ &  $0.6943\pm0.0338$ \\
  	&OWA$_1$ &  $0.6686\pm0.0353$ \\
  	&OWA$_2$ &  $0.6731\pm0.0344$ \\
  	& OWA$_3$ &  $0.6494\pm0.0374$ \\
  	\bottomrule
  \end{tabular}

  \label{tab:four_classes_comparison}

\end{table}

\begin{table}
	\centering
	\caption{\noteb{$P$-values for four classes task comparison for the interval-valued aggregations. $*$ marks statistical differences ($P$-value lesser than 0.05).}}
	\begin{tabular}{llllll}
		\toprule
		 \noteb{Dataset} & {Trad.} &  MD$_2$ & OWA$_1$ &   OWA$_2 $&   OWA$_3$ \\
		\midrule
		%MD$_1$  &  $*$ &  $.22$ &  $.16$ &  $.09$ \\
		\noteb{IV-2a} & MD$_2$  &      &    $*$ &  $*$ &    $*$ \\
		%\hdashline
		& OWA$_1$ &      &        &  $.25$ &  $.25$ \\
		& OWA$_2$ &      &        &        &  $.13$ \\
	\end{tabular}
	
	\begin{tabular}{llllll}
		\toprule
\phantom{IV-2a}&		{MFF} &  MD$_2$ & OWA$_1$ &   OWA$_2$ & OWA$_3$ \\
		\midrule
		%MD$_1$  &  $*$ &  $*$ &    $*$ &  $*$ \\
&		MD$_2$  &      &  $*$ &    $*$ &  $*$ \\
		%\hdashline
&		OWA$_1$ &      &      &  $.25$ &  $*$ \\
&		OWA$_2$ &      &      &        &  $*$ \\
		\bottomrule
	\end{tabular}

%	\caption{\noteb{$P$-values for four classes task comparison for in the interval-valued aggregations. $*$ marks statistical differences ($P$-value lesser than 0.05).}}
	\label{tab:four_classes_p-values}
\end{table}

%\begin{figure}
%  \centering
%%  \includegraphics[width=\linewidth]{gallery/accuracies_full_task_average.pdf}
%  \begin{tabular}{cc}
%	\toprule
%	Aggregation Function & Accuracy \\
%	\midrule
%	OWA$_1$ & 0.6246 \\
%	OWA$_2$ & 0.6318 \\
%	OWA$_3$ & 0.6338 \\
%	MD$_1$ & 0.6424 \\
%	MD$_2$ &  0.6571 \\
%	\bottomrule
%	\end{tabular}
%  \caption{Average accuracy for each different configuration using different aggregation functions in the four classes task.}
%  \label{fig:four_classes_comparison_average}
%\end{figure}
\section{\notea{Comparison with other BCI frameworks}} \label{sec:cmp_others}

\notea{In this Section we have tested the results obtained using the interval-valued MD$_2$ with other non aggregation-based MI BCI frameworks. We have compared our proposal with the results in \cite{8553378} where the authors use multiscale temporal features applied to CSP and Riemannian features; with the works in \cite{hbm23730}, where the authors use two Convolutional Neural Networks (CNNs) to solve different BCI tasks, one with two blocks of Convolution and Image reduction, and the other with four; and with \cite{Lawhern2018} where the authors propose a new Deep Learning architecture with different types of convolutions.}
	
We have computed these comparisons using the same evaluation process as in Section \ref{sec:md_experiments}, with 180 train/test partitions for both datasets. The results are displayed in Table \ref{tab:res_cmp}. \noteb{For the case of the IV-2a dataset}, we have found our results to be inferior to the CSP and Riemannian Multiscale features, and superior to those obtained using CNNs, probably because the CNNs require a much larger number of observations to satisfactory train the model. \noteb{For the IV-2b dataset, the performance of the compared methods is more similar than in the other dataset. In this problem, we found our approach to be the best, closely followed by the CSP multiscale feature and the EEG net.}

\begin{table}
	\centering
	\caption{\noteb{Comparison with non aggregation-based frameworks.}}
	\adjustbox{max width=\linewidth}{
	\begin{tabular}{ccc}
		\toprule
		Framework &  \noteb{IV-2a} & \noteb{IV-2b} \\
		\midrule
		Riemannian Multi Cov. \cite{8553378} & $0.7328\pm0.1325$ & $0.6763\pm 0.1102 $ \\
		CSP Multi Cov. \cite{8553378} & $0.7350\pm0.1507$ & $0.7258\pm 0.1181 $\\
		Shallow CNN \cite{hbm23730} & $0.4862\pm0.1225$ & $0.6996 \pm 0.1387$ \\
		Deep CNN \cite{hbm23730} & $0.3956\pm0.0717$ & $ 0.7045 \pm 0.1259  $\\ %55.59
		EEG Net \cite{Lawhern2018} & $0.5747 \pm 0.1063$ & $0.7229 \pm 0.1281 $\\ %51.16
		\noteb{\L ukasiewicz} $MD_2$ MFF & $0.6943 \pm 0.0338$ & $0.7310\pm 0.0470$ \\
		\noteb{Reinchenbach} $MD_2$ MFF & $0.6905 \pm 0.0328$ & $0.7366\pm 0.0708$ \\
		\bottomrule
	\end{tabular}
}
	\label{tab:res_cmp}
\end{table}

\section{Conclusions} \label{sec:conclusions}
In this work we have presented the interval-valued moderate deviations as a means to aggregate interval-valued data.  We have extended the notion of moderate deviation function to the interval-valued setting, we have analyzed different properties and we have proposed different construction methods. We have, in particular, studied the case \notea{where the width of all the input interval-valued data is the same, and those interval-valued moderate deviation functions which preserve it}.
\noteb{We have applied the interval-valued moderate deviation functions in the decision making phase of two MI BCI frameworks, using fuzzy implication functions to measure the effects of noise in the EEG measurements in each classifier output.} We have studied two different tasks: to discriminate between left hand and right hand classes and among left hand, right hand, tongue and foot classes. We found that the results using interval-valued moderate deviation functions outperform \noteb{the rest of decision making schemes using other numerical and interval-valued aggregations, except for the case of the Choquet integral in the IV-2b dataset using the MFF}.

\noteb{Regarding non aggregation-based BCI frameworks, we found our proposal to beat CNN approaches, but we found our results not as good as the MI BCI framework that used CSP Multiscale Covariance. Since this method focuses on feature extraction, while ours is devoted to improve the decision making phase, we think that combining both approaches can be studied in order to further improve the current results.}

%However, the results are quite different comparing the binary classification and the four classes tasks: the optimal interval-valuate \notea{fuzzy implication function} is different, and so is the best BCI framework for each task.

In our future works we intend to develop moderate deviation based-aggregation functions for $n$-component vectors, and to further explore \notea{the combination of aggregation-based MI BCI frameworks with other MI BCI paradigms}.

\section*{Acknowledgment}
Javier Fumanal Idocin's, Jose Antonio Sanz's, Javier Fernandez's, Harkaitz Goyena's  and Humberto Bustince's research has been supported by the project PID2019-108392GB I00 (AEI/10.13039/501100011033).

Z. Tak\'a\v c acknowledges the support of the grant VEGA 1/0545/20.

\bibliographystyle{IEEEtran}
\bibliography{bci}

\begin{IEEEbiography}[{\includegraphics[width=1in,height=1.25in,clip,keepaspectratio]{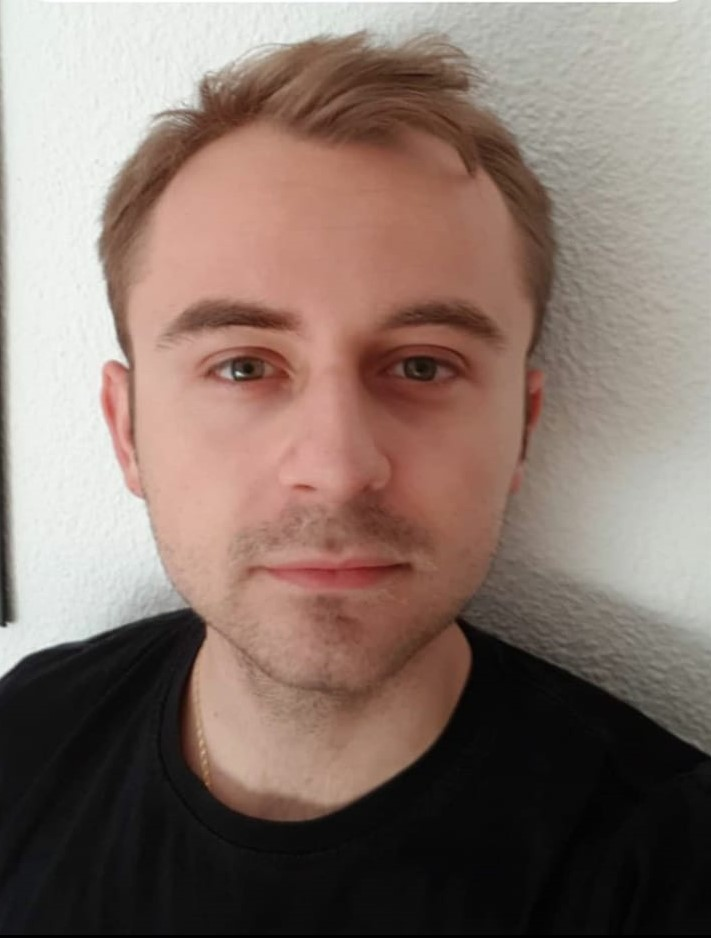}}]{Javier Fumanal Idocin}
	holds a B.Sc in Computer Science at the University of Zaragoza, Spain (2017) and a M.Sc in Data Science and Computer Engineering at the University of Granada, Spain (2018). He is now a PhD Student of the Public University of Navarre, Spain in the department of Statistics, Informatics and Mathematics. His research interests include machine intelligence, fuzzy logic, social networks and Brain-Computer Interfaces.
\end{IEEEbiography}

\begin{IEEEbiography}[{\includegraphics[width=1in,height=1.25in,clip,keepaspectratio]{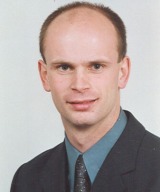}}]{Zdenko Takáč}
	received the Graduate degree in teaching mathematics and physics from the Faculty of Mathematics and Physics, Bratislava, Slovakia, in 1998, and the Ph.D. degree in teaching mathematics with the thesis analysis of mathematical proof from Pavol Jozef afrik University, Koice, Slovakia, in 2007.,Since 1999, he has been a member of the Department of Mathematics, Faculty of Education, Catholic University in Ruomberok, Ruomberok, Slovakia, and since 2010, he has been a member of the Department of Mathematics, Institute of Information Engineering, Automation and Mathematics, Slovak University of Technology in Bratislava, Bratislava, Slovakia. His research interests include uncertainty modeling, fuzzy sets and fuzzy logic, aggregation operators, interval-valued, and type-2 fuzzy sets.
\end{IEEEbiography}

\begin{IEEEbiography}[{\includegraphics[width=1in,height=1.25in,clip,keepaspectratio]{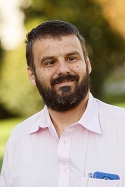}}]{Javier Fernandez}
	received the M.Sc. and Ph.Degrees in Mathematics from the University of Zaragoza, Saragossa, and the University of The Basque Country, Spain, in 1999 and 2003, respectively. He is currently an Associate Lecturer with the Department of Statistics, Computer Science and Mathematics Public University of Navarre, Pamplona, Spain. He is the author or coauthor of approximately 45 original articles and is involved with teaching artificial intelligence and computational mathematics for students of the computer sciences and data science. His research interests include fuzzy techniques for image processing, fuzzy
	sets theory, interval-valued fuzzy sets theory, aggregation functions, fuzzy measures, deep learning, stability, evolution equation, and unique continuation.
\end{IEEEbiography}

\begin{IEEEbiography}[{\includegraphics[width=1in,height=1.25in,clip,keepaspectratio]{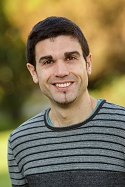}}]{José Antonio Sanz}
	José Antonio Sanz received the M.Sc. and Ph.D. degrees in computer sciences in 2008 and 2011 respectively, both from the Public University of Navarra, Spain. He is currently an assistant professor at the Department of Statistics, Computer Science and Mathematics, Public University of Navarra. He is the author of 36 published original articles in international journals and more than 60 contributions to conferences. He is a member of the European Society for Fuzzy Logic and Technology (EUSFLAT) and the Spanish Association of Artificial Intelligence (AEPIA). He received the best paper award in the FLINS 2012 international conference and the Pepe Millá award in 2014. His research interests include fuzzy techniques for classification problems, interval-valued fuzzy sets, genetic fuzzy systems and medical applications using soft computing techniques.
\end{IEEEbiography}

\begin{IEEEbiography}[{\includegraphics[width=1in,height=1.25in,clip,keepaspectratio]{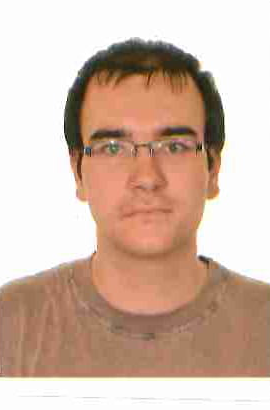}}]{Harkaitz Goyena}
	Harkaitz Goyena received the Graduate degree in Computer Science,
	specializing in Computer Science and Intelligent Systems from the Public
	University of Navarra, Pamplona, Spain, in 2020.
	He is currently studying a master's degree in mathematical investigation
	and modeling, statistics and computing.
\end{IEEEbiography}

\begin{IEEEbiography}[{\includegraphics[width=1in,height=1.25in,clip,keepaspectratio]{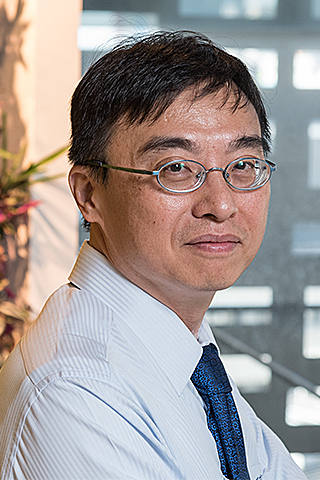}}]{Chin-Teng Lin}
	received the B.S. degree from National Chiao-Tung University (NCTU), Taiwan in 1986, and the Master and Ph.D. degree in electrical engineering from Purdue University, USA in 1989 and 1992, respectively. He is currently the Distinguished Professor of Faculty of Engineering and Information Technology, and Co-Director of Center for Artificial Intelligence, University of Technology Sydney, Australia. He is also invited as Honorary Chair Professor of Electrical and Computer Engineering, NCTU, International Faculty of University of California at San-Diego (UCSD), and Honorary Professorship of University of Nottingham. Dr. Lin was elevated to be an IEEE Fellow for his contributions to biologically inspired information systems in 2005, and was elevated International Fuzzy Systems Association (IFSA) Fellow in 2012. Dr. Lin received the IEEE Fuzzy Systems Pioneer Awards in 2017. He served as the Editor-in-chief of IEEE Transactions on Fuzzy Systems from 2011 to 2016. He also served on the Board of Governors at IEEE Circuits and Systems (CAS) Society in 2005-2008, IEEE Systems, Man, Cybernetics (SMC) Society in 2003-2005, IEEE Computational Intelligence Society in 2008-2010, and Chair of IEEE Taipei Section in 2009-2010. Dr. Lin was the Distinguished Lecturer of IEEE CAS Society from 2003 to 2005 and CIS Society from 2015-2017. He serves as the Chair of IEEE CIS Distinguished Lecturer Program Committee in 2018~2019. He served as the Deputy Editor-in-Chief of IEEE Transactions on Circuits and Systems-II in 2006-2008. Dr. Lin was the Program Chair of IEEE International Conference on Systems, Man, and Cybernetics in 2005 and General Chair of 2011 IEEE International Conference on Fuzzy Systems. Dr. Lin is the coauthor of Neural Fuzzy Systems (Prentice-Hall), and the author of Neural Fuzzy Control Systems with Structure and Parameter Learning (World Scientific). He has published more than 330 journal papers (Total Citation: 22,418, H-index: 67, i10-index: 284) in the areas of neural networks, fuzzy systems, brain computer interface, multimedia information processing, and cognitive neuro-engineering, including about 120 IEEE journal papers.
\end{IEEEbiography}

\begin{IEEEbiography}[{\includegraphics[width=1in,height=1.25in,clip,keepaspectratio]{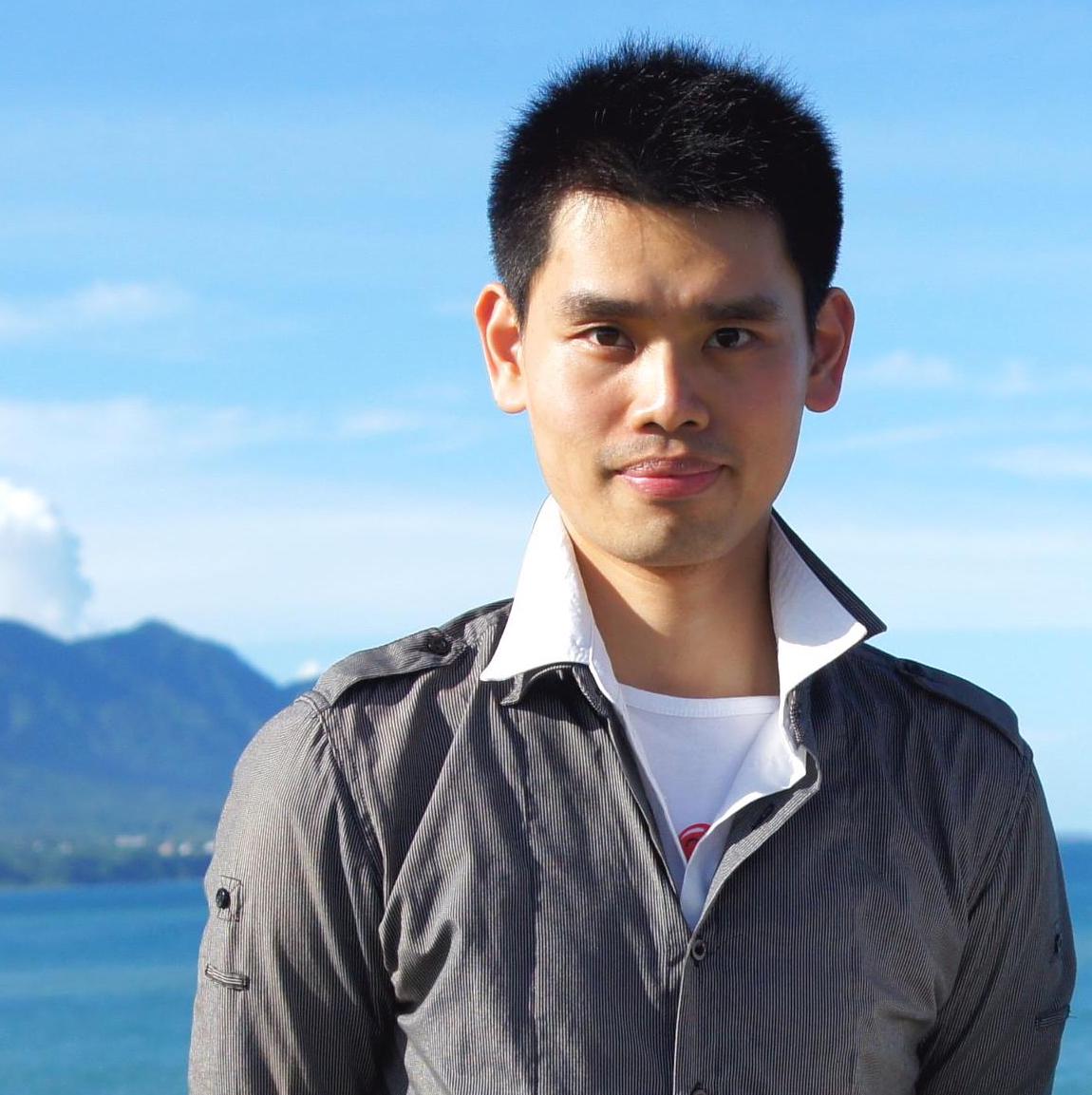}}]{Yu-Kai Wang}
	(M’13) received the B.S. degree in mathematics education from National Taichung University of Education, Taichung, Taiwan, in 2006, the M.S. degree in biomedical engineering from National Chiao Tung University (NCTU), Hsinchu Taiwan, in 2009, and the Ph.D. degree in computer science from NCTU, Hsinchu Taiwan, in 2015. He is currently a Lecturer of Faculty of Engineering and Information Technology, and Co-Director of Motion Platform and Mixed Reality Lab, University of Technology Sydney, Australia. His current research interests include computational neuroscience, human performance modeling, biomedical signal processing, and the brain–computer interface.
\end{IEEEbiography}

\begin{IEEEbiography}[{\includegraphics[width=1in,height=1.25in,clip,keepaspectratio]{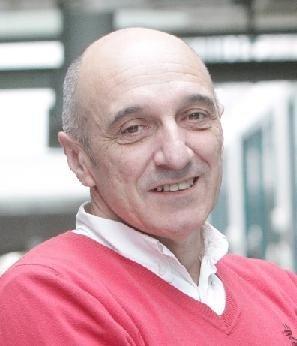}}]{Humberto Bustince}
	received the Graduate degree in physics from the University of Salamanca in 1983 and Ph.D. in mathematics from the Public University of Navarra, Pamplona, Spain, in 1994. He is a Full Professor of Computer Science and Artificial Intelligence in the Public University of Navarra, Pamplona, Spain where he is the main researcher of the Artificial Intelligence and Approximate Reasoning group, whose main research lines are both theoretical (aggregation functions, information and comparison measures, fuzzy sets, and extensions) and applied (image processing, classification, machine learning, data mining, and big data). He has led 11 I+D public-funded research projects, at a national and at a regional level. He is currently the main researcher of a project in the Spanish Science Program and of a scientific network about fuzzy logic and soft computing. He has been in charge of research projects collaborating with private companies. He has taken part in two international research projects. He has authored more than 210 works, according to Web of Science, in conferences and international journals, with around 110 of them in journals of the first quartile of JCR. Moreover, five of these works are also among the highly cited papers of the last ten years, according to Science Essential Indicators of Web of Science. Dr. Bustince is the Editor-in-Chief of the online magazine Mathware \& Soft Computing of the European Society for Fuzzy Logic and technologies and of the Axioms journal. He is an Associated Editor of the IEEE Transactions on Fuzzy Systems Journal and a member of the editorial board of the Journals Fuzzy Sets and Systems, Information Fusion, International Journal of Computational Intelligence Systems and Journal of Intelligent \& Fuzzy Systems. He is the coauthor of a monography about averaging functions and coeditor of several books. He has organized some renowned international conferences such as EUROFUSE 2009 and AGOP. Honorary Professor at the University of Nottingham, National Spanish Computer Science Award in 2019 and EUSFLAT Excellence Research Award in 2019.
\end{IEEEbiography}

\end{document}